\documentclass[fleqn,usenatbib]{mnras}

\usepackage{lineno}
\usepackage[dvipsnames]{xcolor}

\usepackage[T1]{fontenc}

\DeclareRobustCommand{\VAN}[3]{#2}
\let\VANthebibliography\thebibliography
\def\thebibliography{\DeclareRobustCommand{\VAN}[3]{##3}\VANthebibliography}

\usepackage{graphicx}	
\usepackage{amsmath}	
\usepackage{amssymb}	

\newcommand{\angstrom}{\textup{\AA}}

\title[]{Using Host Galaxy Spectroscopy
to Explore 
Systematics in 
the Standardisation of Type Ia Supernovae}

\author[Dixon et al.]{
\parbox{\textwidth}{\Large
M.~Dixon$^1$,
C.~Lidman$^{2,3}$,
J.~Mould$^1$,
L.~Kelsey$^{4,5}$,
D.~Brout$^6$,
A.~M\"oller$^1$,
P.~Wiseman$^5$,
M.~Sullivan$^5$,
L.~Galbany$^{7,8}$,
T.~M.~Davis$^9$,
M.~Vincenzi$^{4,5}$,
D.~Scolnic$^{10}$,
G.~F.~Lewis$^{11}$,
M.~Smith$^5$,
R.~Kessler$^{12,13}$,
A. Duffy$^1$,
E.~N. Taylor$^1$,
C. Flynn$^1$,
T.~M.~C.~Abbott$^{14}$,
M.~Aguena$^{15}$,
S.~Allam$^{16}$,
F.~Andrade-Oliveira$^{17}$,
J.~Annis$^{16}$,
J.~Asorey$^{18}$,
E.~Bertin$^{19,20}$,
S.~Bocquet$^{21}$,
D.~Brooks$^{22}$,
D.~L.~Burke$^{23,24}$,
A.~Carnero~Rosell$^{15,25,26}$,
D.~Carollo$^{27}$,
M.~Carrasco~Kind$^{28,29}$,
J.~Carretero$^{30}$,
M.~Costanzi$^{27,31,32}$,
L.~N.~da Costa$^{15}$,
M.~E.~S.~Pereira$^{33}$,
P.~Doel$^{22}$,
S.~Everett$^{34}$,
I.~Ferrero$^{35}$,
B.~Flaugher$^{16}$,
D.~Friedel$^{28}$,
J.~Frieman$^{13,16}$,
J.~Garc\'ia-Bellido$^{36}$,
M.~Gatti$^{37}$,
D.~W.~Gerdes$^{17,38}$,
K.~Glazebrook$^1$,
D.~Gruen$^{21}$,
J.~Gschwend$^{15,39}$,
G.~Gutierrez$^{16}$,
S.~R.~Hinton$^9$,
D.~L.~Hollowood$^{40}$,
K.~Honscheid$^{41,42}$,
D.~Huterer$^{17}$,
D.~J.~James$^6$,
K.~Kuehn$^{43,44}$,
N.~Kuropatkin$^{16}$,
U.~Malik$^3$,
M.~March$^{37}$,
F.~Menanteau$^{28,29}$,
R.~Miquel$^{30,45}$,
R.~Morgan$^{46}$,
B.~Nichol$^{47}$,
R.~L.~C.~Ogando$^{39}$,
A.~Palmese$^{48}$,
F.~Paz-Chinch\'{o}n$^{28,49}$,
A.~Pieres$^{15,39}$,
A.~A.~Plazas~Malag\'on$^{50}$,
M.~Rodriguez-Monroy$^{18}$,
A.~K.~Romer$^{51}$,
E.~Sanchez$^{18}$,
V.~Scarpine$^{16}$,
I.~Sevilla-Noarbe$^{18}$,
M.~Soares-Santos$^{17}$,
E.~Suchyta$^{52}$,
G.~Tarle$^{17}$,
C.~To$^{41}$,
B.~E.~Tucker$^3$,
D.~L.~Tucker$^{16}$,
T.~N.~Varga$^{53,54,55}$}}

\date{Accepted XXX. Received YYY; in original form ZZZ}

\pubyear{2022}

\begin{document}
\label{firstpage}
\pagerange{\pageref{firstpage}--\pageref{lastpage}}
\maketitle


\begin{abstract}


We use stacked spectra of the host galaxies of photometrically identified type Ia supernovae 
(SNe Ia) from the Dark Energy Survey (DES) to search for correlations between Hubble diagram residuals and the spectral properties of the host galaxies. Utilising full spectrum fitting techniques on stacked spectra binned by Hubble residual, we find no evidence for trends between Hubble residuals and properties of the host galaxies that rely on spectral absorption features ($< 1.3\sigma$), such as stellar population age, metallicity, and mass-to-light ratio. However, we find significant trends between the Hubble residuals and the strengths of [OII] ($4.4\sigma$) and the Balmer emission lines ($3\sigma$). These trends are weaker than the  well known trend between Hubble residuals and host galaxy stellar mass ($7.2\sigma$) that is derived from broad band photometry. After light curve corrections, we see fainter SNe Ia residing in galaxies with larger line strengths. We also find a trend (3$\sigma$) between Hubble residual and the Balmer decrement (a measure of reddening by dust) using H${\beta}$ and H${\gamma}$. The trend, quantified by correlation coefficients, is slightly more significant in the redder SNe Ia, suggesting that bluer SNe Ia are relatively unaffected by dust in the interstellar medium of the host and that dust contributes to current Hubble diagram scatter impacting the measurement of cosmological parameters.

\end{abstract}




\section{Introduction}
Standardised Type Ia supernovae (SNe Ia) are important distance indicators used to probe the expansion history of the universe (\citealp{Reiss1998,1999perl}) and measure the local Hubble constant ($H_{0}$) \citep{Dhawan_2020,Khetan2020, freedman2021, 2022Riess}.
SNe Ia are extremely bright stellar explosions with a peak magnitude in the B-band of $\sim -19.5$ mag \citep{Childress_2017}. They are standardised through light curve corrections that account for the relationship between the peak magnitude and the stretch or width of the SN Ia light curve \citep{1993ApJ...413L.105P} and the relationship between the peak magnitude and SN Ia colour \citep{1998A&A...331..815T}. 

Many studies over the last two decades have further improved the standardisation of SNe Ia by making use of broad band photometry and spectroscopy of host galaxies to correct for trends between the colour and stretch corrected luminosities of SNe Ia and the properties of the host. These include gas phase metallicity \citep{2011d'andrea,Pan2013}, stellar age \citep{Childress_2013,2019rose}, specific star-formation rate \citep{Lampeitl_2010, 2011d'andrea, Childress_2013, Rigault_2020} and  rest-frame colour \citep{2018roman, 2021Kelsey, kelsey2022}.
The most commonly used correction is with the stellar mass of the host galaxy. The correction is applied as a step function around $10^{10}M_{\bigodot}$, and is referred to in the literature as the "mass step" where SNe Ia in high mass galaxies ($>10^{10}M_{\bigodot}$) are more luminous after correction than SNe Ia in low mass galaxies (\citealp{kelly2010,Sullivan_2010,G10,Childress_2013,uddin_mould_lidman_ruhlmann-kleider_hardin_2017,Smith_2020,2021Kelsey}).

The mass step has been found with varying significance. A mass step of $0.039\pm0.016$ mag was uncovered using the Pantheon sample \citep{2018Scolnic}, while \cite{Brout_2019} found no evidence of a mass step ($0.009\pm0.018$ mag), when using SNe Ia from the Dark Energy Survey (DES). However, \cite{Smith_2020} observed a significant mass step of $0.040\pm0.019$ mag when utilising the same sample, and showed that the magnitude of the mass step was affected by the way survey selection biases were computed. The size of the mass step also depends on the method used to standardise SN Ia luminosities. \citet{Boone2021b} find a mass step of $0.092\pm0.026$ mag using their SN Ia sample and the SALT2 standardisation method \citep{Betoule2014}, and a reduced step of $0.040\pm0.020$ mag when using an alternative standardisation method (Twin Embedding Model).


The relationship between SN color and peak brightness is modelled with a single term in the Tripp equation \citep{1998A&A...331..815T}. However, it is likely that there are a least two mechanisms at play: an intrinsic relation between luminosity and color and reddening and attenuation by dust. There is no reason why these two mechanisms should follow the same relationship. Attenuation by dust is wavelength dependent and is defined by an attenuation law. The law depends on the dust grain size and composition, and varies within the Milky Way and from galaxy to galaxy. It has been found to correlate with host galaxy properties. For example, high mass galaxies tend to have greater attenuation in the V-band and shallower attenuation laws\footnote{Attenuation laws are often parameterised with $R_V$, the total to selective extinction ratio. Shallower attenuation laws have higher $R_V$ values.}, when compared with low mass galaxies \citep{2018Salim}.

In a recent work, \citet{brout2021dust} found that the mass step could be reproduced in a model that allowed the attenuation law and the amount of attenuation to vary with host galaxy mass. Furthermore, the observed trends in Hubble residuals and SN color were well captured by a model (hereafter referred to as the BS21 model) which combined an intrinsic SN colour-magnitude relation with attenuation laws that depended on host galaxy mass.
Independently, \cite{2021johansson} recovered a significant mass step when assuming a fixed attenuation law with $R_V=2$. However, when utilising an individual best fit value of $R_{V}$ for each SN Ia, the mass step vanished. Additionally, \cite{meldorf2022} find agreement between the fitted colour slope from the SN Ia photometry and fitted host galaxy  $R_{V}$ from host photometry when splitting the sample on galaxy population.


The goal of our work is to
search for correlations between Hubble residuals and SN Ia host galaxy properties. Instead of deriving host galaxy properties from broad band photometry as most studies have done (e.g. \citealp{Sullivan_2010, Rigault_2013, uddin_mould_lidman_ruhlmann-kleider_hardin_2017, 2019rose, Smith_2020, 2021Kelsey, 2021johansson}), we use spectra. Spectra offer the possibility of capturing trends that cannot be extracted using photometry. Using spectra from the Sloan Digital Sky Survey (SDSS),  \cite{2013JohanssonJ} find no significant trends between properties measured from spectra and Hubble residuals. Using a larger sample of SDSS SNe Ia, \citet{2016campbell} find a weak trend between Hubble residuals and galaxy metallicity, while \citet{2021Galbany} recently found trends between Hubble residuals with metallicity, specific-star formation rate and the equivalent width of H$\alpha$.


The sections of this paper consist of the following:
In Section \ref{section2}, we describe the SN Ia sample used in the analysis. In Section \ref{section3}, we  describe the spectral fitting technique that is used to extract galaxy properties from stacked host galaxy spectra. Then, in Section \ref{section4} we search for trends between Hubble residuals and these properties. Section \ref{section5} discusses the results.

\section{Surveys and Data Selection}
\label{section2}
\subsection{The Dark Energy Survey (DES)}

The Dark Energy Survey (DES) was designed to constrain the properties of dark energy using four probes: galaxy clusters, weak lensing, Baryon Acoustic Oscillations, and SNe Ia \citep{Abbott2016}. Running for six observing seasons, from 2013 to 2019, and using the 570 megapixel Dark Energy Camera (DECam; \citealp{Flaugher2015}) mounted on the 4-m Victor M. Blanco  Telescope \citep{Abbott2016}, DES consists of two surveys: a wide survey covering 5100 square degrees; and a 27 square degree time domain survey covering 10 deep fields with weekly cadence in the griz bands. DES has catalogued hundreds of millions of galaxies and has discovered thousands of supernovae \citep{hartley2020dark}. 
The data is passed through the DES Image Processing Pipeline \citep{Morganson_2018}, where transients are identified using a difference imaging pipeline \citep{2019BroutSMP}.

\subsection{The Australian Dark Energy Survey (OzDES)}

The Australian Dark Energy Survey (OzDES) was undertaken over the same six observing seasons as DES \citep{Yuan2015, Childress_2017,Lidman_2020}. Observations were performed on the 3.9-m Anglo-Australian Telescope (AAT) at Siding Spring Observatory and utilised the AAOmega spectrograph with the 2dF field positioner \citep{Yuan2015} to target the 10 deep fields that are a part of the DES time domain survey. The 2dF fibre positioner  can place up to 8 guide fibres and 392 science fibres within a 2.1 degree field, matching the field of view of the DECam imager. Its primary aims were measuring redshifts of SN hosts, confirming the spectral type of SNe, and monitoring AGNs over a wide range of redshift \citep{hoorman2019}. OzDES spectra have previously been used in a number of studies (\citealp{Wiseman2020b, Pursiainen2020, Smith2020b}). The faintest objects have an apparent magnitude  $r\approx$ 24 mag and the wavelength coverage is between 3700 and 8800$\angstrom$.  The second OzDES data release includes 375,000 spectra of 39,000 objects. A full description of the data can be found in \citet{Lidman_2020}. 

\subsection{Type Ia Supernovae Sample}
We utilise the DES5YR-SN photometric sample described in \cite{2022Anais}. The SNe Ia were classified with the SuperNNova (SNN) classifier \citep{2020Anais}. SNN identifies SNe Ia with high accuracy and high purity. The sample contains 1484 photometrically classified SNe Ia, and includes light curve quality cuts for the colour ($c$) and stretch ($x1$): $-0.3 < c < 0.3, -3 < x1 < 3$ \citep{Betoule2014}.

We then implement specific cuts (Table \ref{Table:cuts}). 
For each DES transient, the host is identified as the galaxy with the smallest directional light radius (DLR; \citealp{Gupta_2016}) from the deep image stacks described in \citet{Wiseman_2020}. We then match the host with objects in the OzDES redshift catalogue \citep{Lidman_2020}, and a cut is made regarding the quality of  the spectroscopic redshift.
The OzDES redshift reliability criterion is determined by the quality of the matched spectroscopic features. A flag = 3 redshift is based on identifying a single strong spectroscopic feature or multiple weak features, and is likely to be correct more than 95$\%$ of the time. Redshifts with a flag = 4 are based on multiple strong features and are likely to be correct more than 99$\%$ of the time. We require the quality flag $\geq$ 3 in our analysis, and omit any galaxies without a spectroscopic redshift.

We also only use spectra of the hosts that are not contaminated by supernova light. For this reason we select OzDES host galaxies that were observed two months before or five months after the SN Ia peak luminosity occurred. Around maximum brightness, light from the SN can contaminate and potentially bias the spectrum of the host. These cuts reduce the number of SNe Ia in our sample from 1484 to 625.

\begin{table}
\centering
\caption{The size of the sample after each cut is applied.}
\begin{tabular}{lr}
\hline
Cut & Number of SNe Ia \\ 
\hline
SN Ia Photometric Sample & 1484 \\ 
Matching SN transient with a host galaxy \\ with an OzDES observation & 1181 \\ Redshift reliability $>95\%$ & 874 \\
OzDES spectra not contaminated \\ by SN light & 625 \\
\hline
\end{tabular}
\label{Table:cuts}
\end{table}

\section{Methodology} \label{section3}

\subsection{Hubble Residuals}
Hubble residuals ($\Delta\mu$) are the deviation between the inferred distance modulus ($\mu_\mathrm{obs}$) and the expected value at a given redshift and cosmological model ($\mu_\mathrm{theory}(z)$):
\begin{equation}
    \Delta \mu = \mu_\mathrm{obs} - \mu_\mathrm{theory}(z).
\end{equation}
A positive (negative) Hubble residual implies that the SN is fainter (brighter) than the model. Each SN Ia light curve is fit using the SALT2 light curve model \citep{G10} that was trained for the Joint Light Curve Analysis \citep{Betoule2014}. This fit gives a colour ($c$), stretch ($x_{1}$), and apparent magnitude at peak brightness in the B-band ($m_{B}$). These parameters are then used to estimate the distance modulus ($\mathrm{\mu}_\mathrm{obs}$), by applying the modified Tripp formula \citep{1998A&A...331..815T}:
\begin{equation}
    \mathrm{\mu}_\mathrm{obs} = \mathrm{m}_\mathrm{B} + \alpha x_{1} - \beta c + M_{0} + \gamma G_{host} + \mu_{bias}.
\end{equation}
The nuisance parameters ($\alpha$, $\beta$) are set to $\alpha = 0.156 \pm 0.012$ and $\beta = 3.201 \pm 0.131$ \citep{Smith_2020} and the absolute magnitude is set to $M_{0} = -19.5$. The bias correction term, $\mu_{bias}$, is obtained from simulations using BBC \citep{Kessler_2017} and takes into account survey selection effects. 

The term $\gamma G_\mathrm{host}$ is the correction for the mass step and is usually formulated as:

\begin{equation}
    G_\mathrm{host} = \left\{ \begin{array}{cl}
+1/2 & \mathrm{for} \ M_{*} > M_\mathrm{step} \\
-1/2 & \mathrm{otherwise},
\end{array} \right.
\end{equation}
where $M_{*}$ is the stellar mass of the SN host galaxy, and $\gamma$ is the size of the mass step. The division point ($M_\mathrm{step}$) is generally taken as $10^{10}$ $\mathrm{M_{\bigodot}}$. 

We exclude the mass step term ($\gamma G_\mathrm{host}$) and use a 1D bias correction \citep{Smith_2020} in our analysis.  This allows us to investigate the significance of any dependencies between Hubble residuals and host galaxy properties in our sample.

\begin{figure}
    \centering
    \includegraphics[scale=0.155]{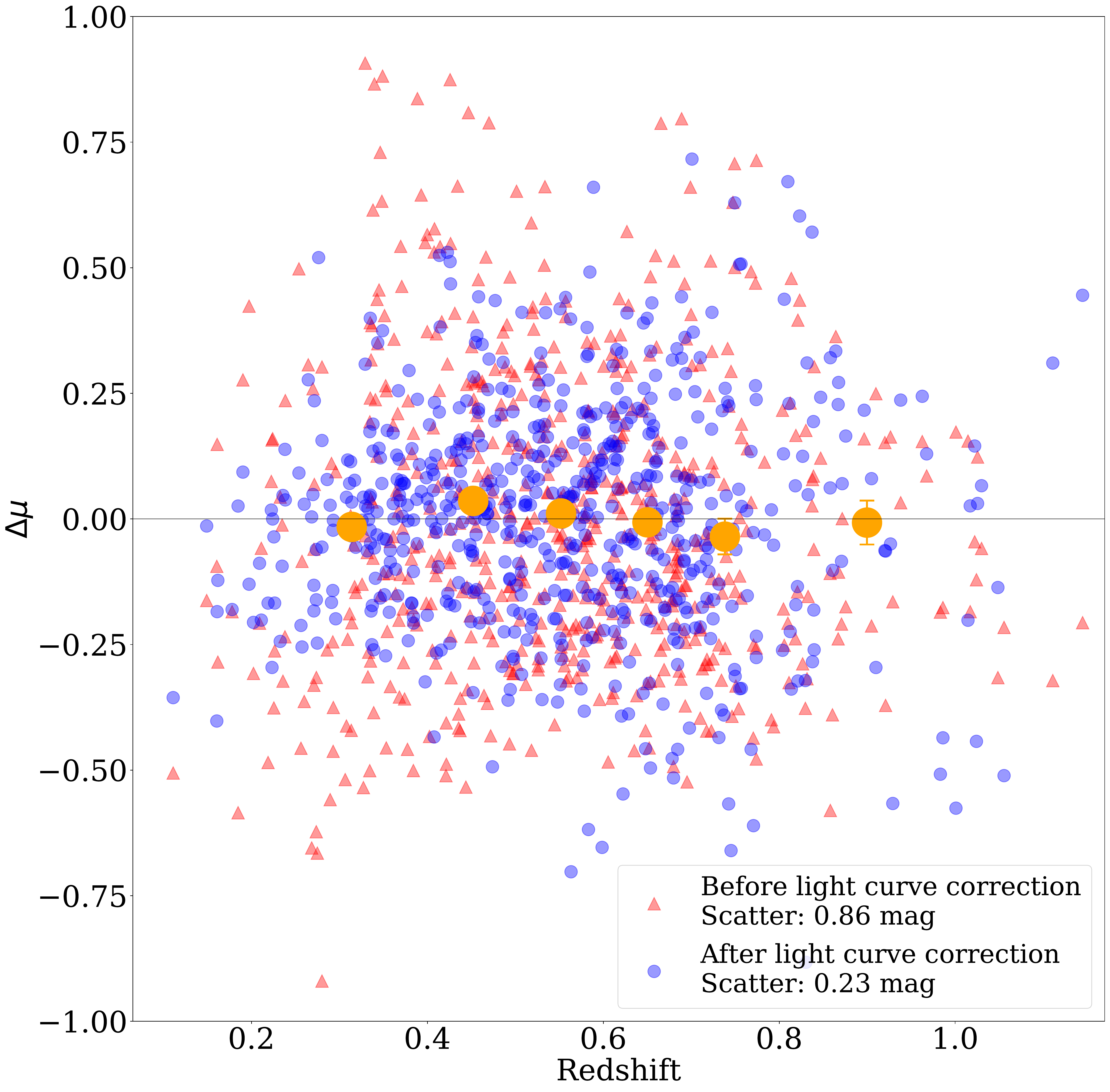}
    \caption{Hubble residual diagram for the OzDES SN Ia host galaxy sample containing 625 SNe Ia. The red points are the residuals before light curve correction and the blue points have been corrected for colour, stretch and survey selection biases. The median redshift of the sample is $z \sim$ 0.55. Mean scatter of the residuals split into redshift bins are shown as the orange points.}
        
    
    \label{fig:HR_YR5}
\end{figure}

Figure \ref{fig:HR_YR5} shows the Hubble residual diagram with the 625 SNe Ia remaining after the cuts to the DES5YR photometrically selected SN Ia sample. For each SN Ia we plot $\Delta\mu$ and then apply the light curve and 1D bias corrections. The scatter reduces from 0.86 to 0.23 mag after these corrections. 

\begin{figure} 
    \centering
    \includegraphics[scale=0.24]{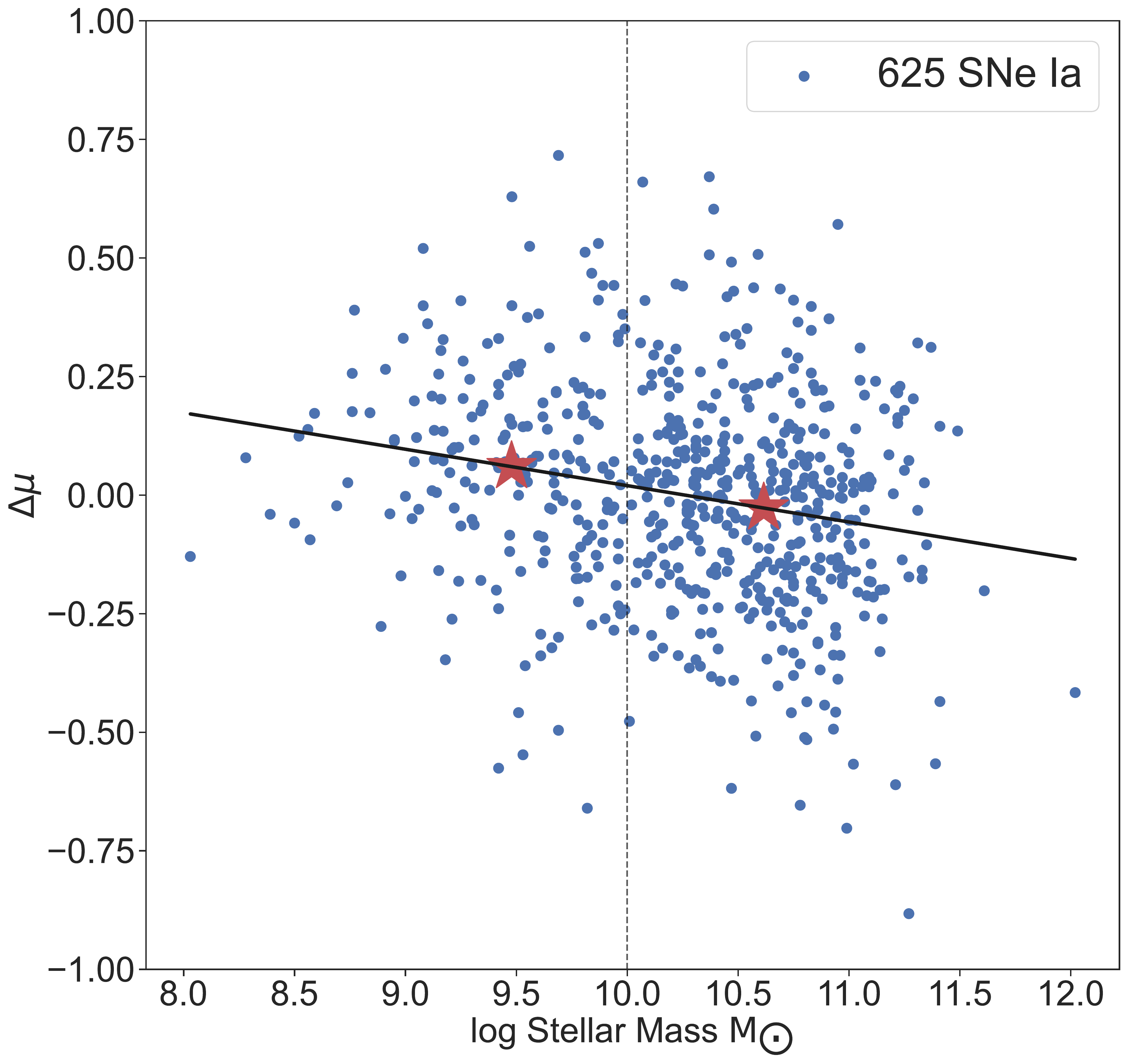}
    \caption{The DES5YR photometrically selected SN Ia  sample after selection cuts (Table \ref{Table:cuts}), standard light curve corrections ($\alpha = 0.156$ and $\beta = 3.201$) and 1D bias corrections, showing Hubble residual against host galaxy stellar mass. The black line corresponds to a linear fit and the red stars are the average Hubble residual either side of $10^{10} \mathrm{M_{\bigodot}}$. The linear fit has a slope $-0.087 \pm 0.014$ and the mass step between the averaged points is $0.087 \pm 0.011$, which is comparable to that found for DES3YR SN Ia sample. \citep{Smith_2020}.}
    \label{fig:mass}
\end{figure}

Figure \ref{fig:mass} shows the trend between host stellar mass and Hubble residual. We make use of masses obtained using the method described in \cite{Smith_2020}, which were estimated using broadband photometry of the host galaxies in the deep image stacks of the DES-SN fields \citep{Wiseman_2020}. The trend is fitted with a straight line, and as seen in other large SN surveys, the more luminous SNe Ia  after light curve correction (negative residuals) reside in galaxies with higher stellar masses. The median stellar mass of the host galaxies is $10^{10.34} M_{\bigodot}$, so there are more galaxies with masses above  $10^{10} M_{\bigodot}$, the canonical mass used for the mass step, than below it.

\subsection{Coadding and Stacking OzDES Host Galaxy Spectra}
\label{Coadding}

In this work we use the term coadding when combining spectra of the same object and stacking when combining spectra of different objects.

\subsubsection{Coadded OzDES Spectra} 

The SN host galaxies in our sample cover a broad range of apparent magnitude, between $18 < r < 24$. Integration times for the faintest hosts were very long (up to 100 hours in some cases), and most targets were observed over several nights. During this time, the seeing and transparency can vary greatly, which results in large fluctuations in the amount of light entering the fibres. Consequently, the throughput varied by an order of magnitude. To overcome this variability, the OzDES data reduction pipeline scales the data with the inverse of the median flux of the extracted spectrum.\footnote{In detail, the pipeline scales each spectrum with either the median of the flux or 0.1 times the square root of the median variance, whichever is greater. The scaling is computed from data obtained in the red arm of AAOmega and is applied to both the red and blue arms.} The variance spectrum is scaled by the square of this scaling factor. The coadded spectrum is the variance weighted average of the scaled spectra. By design, the continuum of the red half of the coadded spectrum has a median that is close to one. 

\subsubsection{Stacking OzDES Spectra} 

Due to the low signal-to-noise ratio (S/N) of coadded spectra, we split the SN Ia host galaxies into stacks based on the size of the Hubble residual and compute the variance-weighted average spectrum for each stack. The first step is to deredshift each spectrum. This is done using linear interpolation to a common wavelength grid.

The spectra are stacked using the same procedure as that used for the individual spectra. Again, the stacked spectra have a median that is close to one, as can be seen in Figure 3. 

To understand how uncertainties affect our analysis, we implement two approaches. We capture the formal variance by perturbing each point in the averaged spectrum. The magnitude of the perturbation is drawn from a Gaussian that has a mean of zero and a variance that is determined from the variance spectrum. This is repeated 1000 times. However, the formal variance does not capture the intrinsic variability of the individual spectra. To capture the intrinsic scatter we perform 1000 bootstrap samples of the spectra used in the stack.

\begin{figure*}
    \centering
    \includegraphics[scale=0.18]{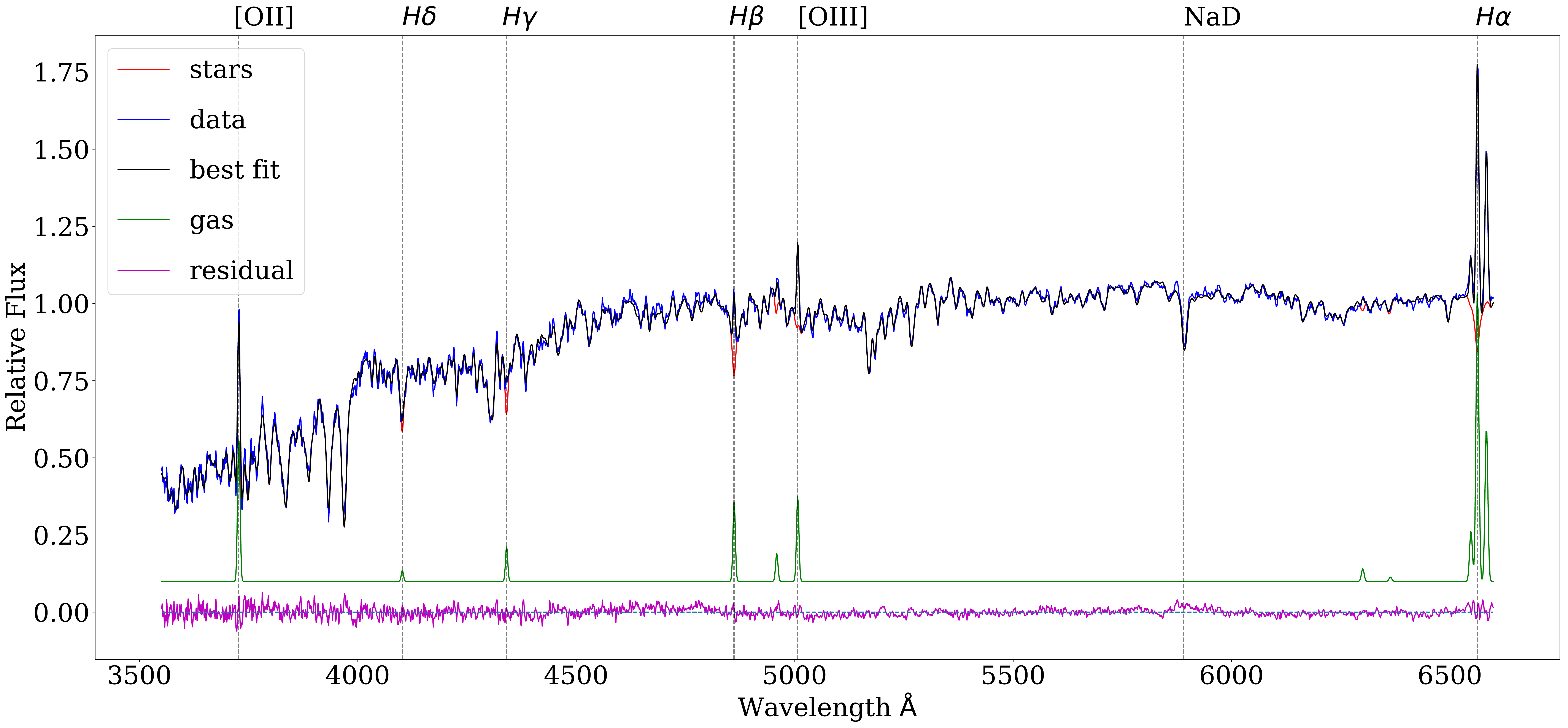}
    \caption{An example of running pPXF on a OzDES host galaxy stack (76 objects with $z \leq 0.35$). The blue line represents the input spectrum, the black line is the best fit, the red and green lines (the latter shifted up 0.1 for clarity) are the stellar and emission line components, respectively, and the purple line is the residual flux. Note how the Balmer emission lines fill in the absorption lines from the stars.}
    \label{fig:ppxf}
\end{figure*}

\begin{table*}
\caption{pPXF output for each of the stacked spectra, where the uncertainties were obtained using bootstrap resampling.}
\begin{tabular}{cccccc}
\hline
$\Delta \mu$ & log Stellar Age (yrs) & Mass-to-light ratio & Metallicity [M/H] & Equivalent Width [OII] ($\angstrom$) & log Stellar Mass $\mathrm{M_{\bigodot}}$ \\
\hline
$-0.37$ & $ 9.68   \pm 0.07 $ & $ 2.12 \pm 0.22 $ & $ -0.13 \pm 0.02 $ & $ -5.81 \pm 0.73 $ & $ 10.46   \pm 0.09 $ \\
$-0.23$ & $ 9.79 \pm 0.07 $ & $ 2.47 \pm 0.35 $ & $ -0.2 \pm 0.04 $ & $ -6.26 \pm 1.03 $ & $ 10.53 \pm 0.06 $ \\
$-0.17$ & $ 9.8 \pm 0.07 $ & $ 2.53 \pm 0.36 $ & $ -0.2 \pm 0.04 $ & $ -5.78 \pm 1.23 $ & $ 10.40 \pm 0.09 $ \\
$-0.12$ & $ 9.88 \pm 0.09 $ & $ 2.95 \pm 0.65 $ & $ -0.24 \pm 0.07 $ & $ -6.69 \pm 2.38 $ & $ 10.31 \pm 0.09 $ \\
$-0.06$ & $ 9.83 \pm 0.05 $ & $ 2.35 \pm 0.24 $ & $ -0.51 \pm 0.12 $ & $ -8.46 \pm 1.24 $ & $ 10.26 \pm 0.09 $ \\
$-0.01$ & $ 9.73 \pm 0.09 $ & $ 2.10 \pm 0.30 $ & $ -0.23 \pm 0.06 $ & $ -7.2 \pm 1.35 $ & $ 10.27 \pm 0.09 $ \\
0.02 & $ 9.87 \pm 0.05 $ & $ 2.87 \pm 0.39 $ & $ -0.21 \pm 0.03 $ & $ -8.7 \pm 1.74 $ & $ 10.35 \pm 0.09 $ \\
0.06 & $ 9.81 \pm 0.08 $ & $ 2.13 \pm 0.32 $ & $ -0.47 \pm 0.17 $ & $ -14.38 \pm 1.5 $ & $ 10.09 \pm 0.08 $ \\
0.10 & $ 9.48 \pm 0.18 $ & $ 1.45 \pm 0.34 $ & $ -0.13 \pm 0.09 $ & $ -10.53 \pm 1.84 $ & $ 10.02 \pm 0.10 $ \\
0.16 & $ 9.68 \pm 0.11 $ & $ 1.88 \pm 0.50 $ & $ -0.42 \pm 0.2 $ & $ -9.21 \pm 2.72 $ & $ 10.12 \pm 0.11 $ \\
0.23 & $ 9.89 \pm 0.05 $ & $ 2.79 \pm 0.34 $ & $ -0.33 \pm 0.15 $ & $ -9.38 \pm 2.29 $ & $ 10.09 \pm 0.10 $ \\
0.39 & $ 9.81 \pm 0.05 $ & $ 2.47 \pm 0.28 $ & $ -0.18 \pm 0.02 $ & $ -10.66 \pm 2.11 $ & $ 10.02 \pm 0.09 $ \\
\hline
\end{tabular}
\label{table:ppxf_output_table}
\end{table*}

The method of coadding and stacking spectra has a number of consequences. The integrated line fluxes are scaled relative to the continuum between 5700\AA\ and 8600\AA, so line fluxes are not absolute. Instead, they can be better described as continuum normalised fluxes, which we simply refer to as line strengths throughout this paper. Line flux ratios are unchanged as the same scaling is applied to all lines. Equivalent widths are also unchanged as the same scaling is applied to both the continuum and the line.

\subsubsection{Systematic uncertainties in the calibration of OzDES data}

The conversion from counts measured on the detector to spectral flux
densities requires two sensitivity curves, one for each arm of the
AAOmega spectrograph. To determine these curves, OzDES allocated 8 to
12 fibres to F stars in every configuration. These stars were observed at
the same time as the galaxies. In principle, OzDES could have
produced sensitivity curves for every observation, but chose
instead to use curves that were determined using data obtained during
the first two OzDES observing seasons.

When done in this way, the spectral flux densities are relative and
not absolute, as flux losses caused by the finite area of the fibres
are not taken into account. Additionally, other errors can lead to
chromatic warping of the spectra. These errors include field and
wavelength dependent aberrations from the 2dF corrector, errors in the
positioning of the fibres, and variations in the natural seeing. As
demonstrated in \cite{hoorman2019}, the warping of the spectra can
be removed by applying a second order polynomial that is derived from
the comparison of broad band photometry of the objects with synthetic
photometry derived from the flux calibrated spectra. This works well
for relatively bright sources, but works less well for sources
considered here, many of which are considerably fainter than the
sky. For such objects, a small error in the subtraction of sky can
lead to a big impact on the continuum and to an erroneous estimate of
the amount of warping required.

Hence, the spectra used in this work are stacked without warping. The systematic error associated with neglecting the
correction is small and can be estimated from the F stars
themselves. As noted above, about a dozen F stars are allocated fibres and observed at the same time as the supernovae host
galaxies. Over the 6 years OzDES ran, nearly 10,000 F star
spectra were obtained. We use these spectra to estimate how much tilt needs to be
applied so that the observed and modeled spectra of the the F stars
agree. We find a scatter of 24\% per 100 nm for the tilt in the blue
arm and a scatter of 7\% per 100 nm for the tilt in the red arm. The
scatter is for a single observation. The stacked host galaxy spectra
consist of spectra from around 100 host galaxies that are randomly
distributed across the 2dF field. Under the assumption that the tilts
are also randomly distributed, the tilt in the stacked spectra will
be lower, by approximately the tilt in the spectrum of a single galaxy
divided by the square root of the number of galaxies. The tilt in the
combined spectra is expected to be of the order of a few percent per
100 nm. It is considerably smaller than the statistical uncertainties
and hence is not included in our analysis.

\begin{figure*}
    \centering
    \includegraphics[scale=0.17]{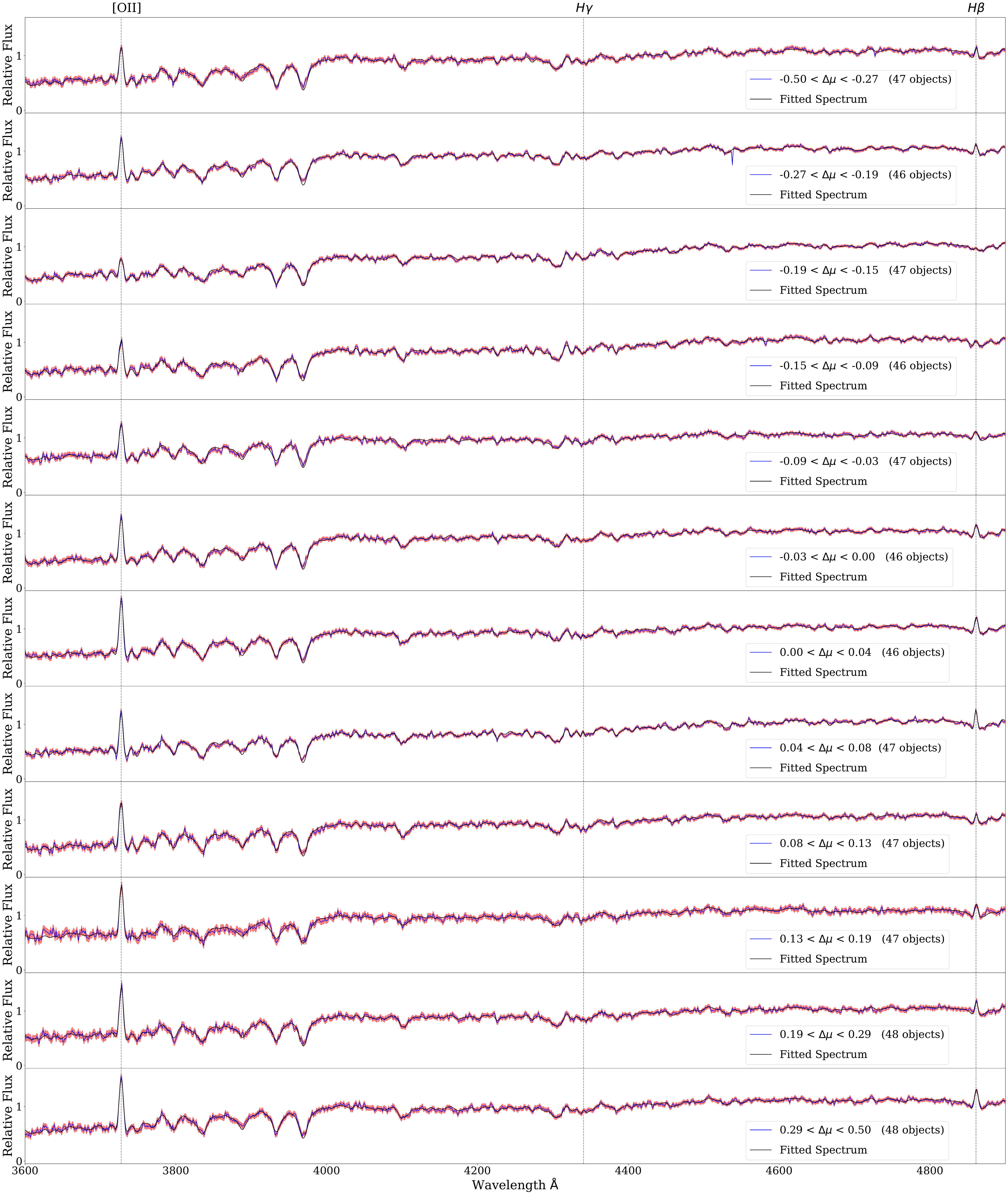}  \caption{Stacked OzDES spectra which are binned by Hubble residual for $z \leq 0.8$. Each stack contains a similar number of objects, resulting in high S/N. The blue line represents the data while the black line is the fit obtained using pPXF. The filled red area indicates the standard deviation obtained from bootstrap resampling.}
    \label{fig:stacks}
\end{figure*}

\begin{figure*}
    \centering 
    \includegraphics[scale=0.17]{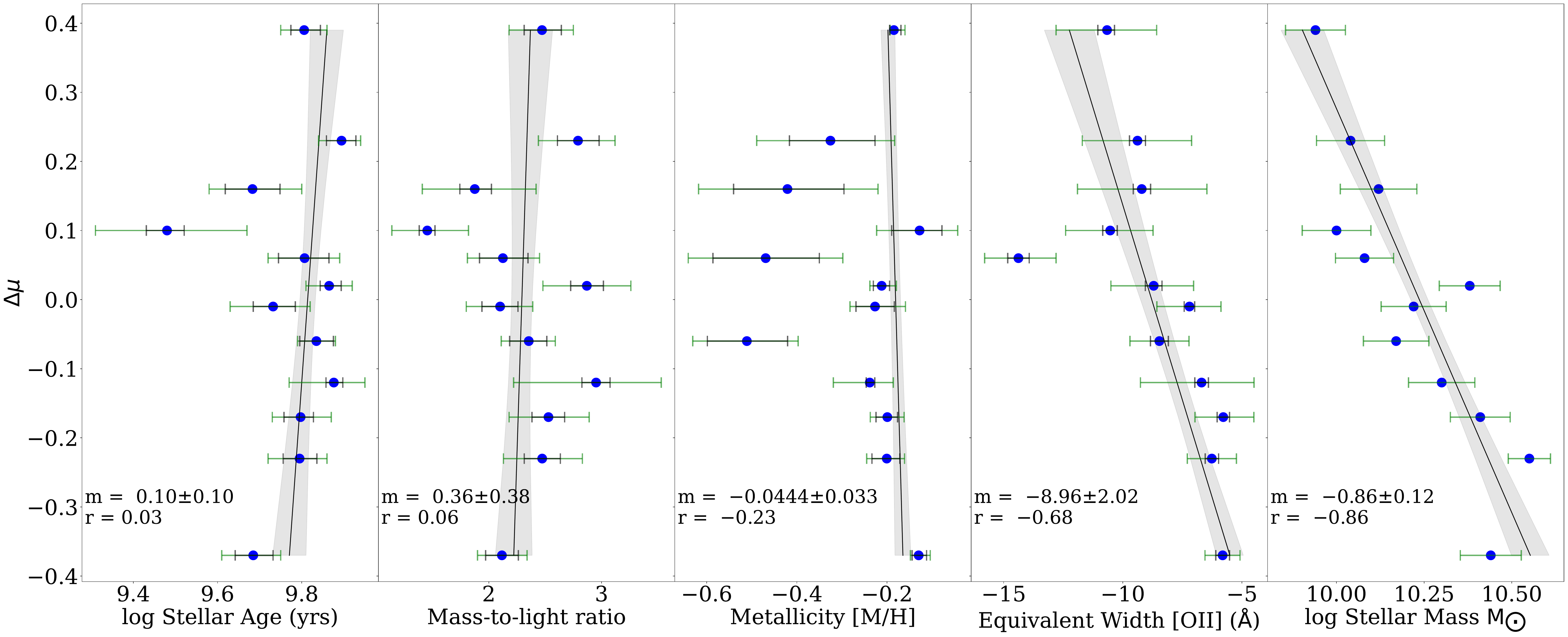}
    \caption{Left to right: Trends between host galaxy properties and Hubble residuals. The first four plots are derived using spectra, while the last plot is derived using photometry. The outer green errorbars  are computed by bootstrap resampling and the inner black errorbars are obtained from the formal variance. We show a linear fit for each property, with 1$\sigma$ confidence intervals. $m$ represents the slope. The trend between host stellar mass and Hubble residual is the most significant and has the highest Pearson correlation coefficient, r.}
    \label{fig:finer_hr_binning}
\end{figure*}

\begin{figure}
    \centering
    \includegraphics[scale=0.175]{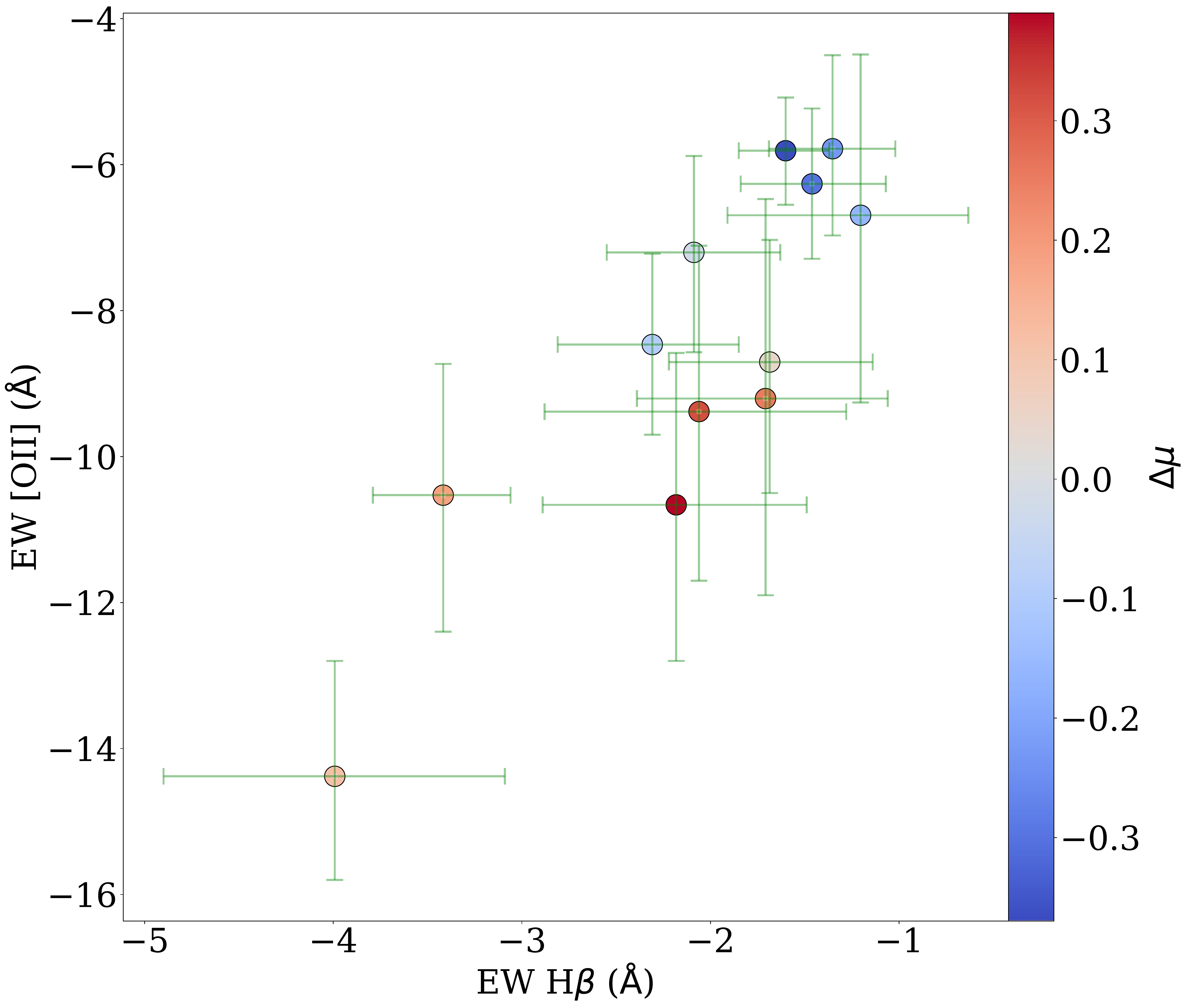}
    \caption{A plot comparing the EWs of [OII] and  H$\beta$. Note how the EWs of both lines  correlate with Hubble residual. Hosts with the strongest emission lines (more negative EWs) also tend to host SN Ia with positive Hubble residuals (colour bar).}
    \label{fig:EW}
\end{figure}

\subsection{Spectral Fitting}
Penalized Pixel-Fitting (pPXF) is used to extract information on the stellar population from our stacked galaxy spectra. pPXF is a full spectrum fitting technique which utilises a penalised maximum-likelihood approach (\citealp{2004_Cappellari}). pPXF is one option that is widely used and trusted in its full spectral fitting implementation. While other useful tools such as Prospector \citep{2021_prospector} could be utilised, we find the excellence of the fit provided by pPXF sufficient for our approach in deriving host properties to search for trends with Hubble residual.

A single stellar population (SSP) model consists of groups of stars with the same metallicity, a given IMF and all formed at the same time. A general stellar population is then a linear combination of SSPs. The MILES spectral library is used to build the synthetic spectrum, ranging in age from 0.63 to 17.8 Gyr and metallicity from $-2.32$ to 0.22 [M/H] with a unimodal IMF with a slope of 1.30 \citep{miles}. They were obtained at the 2.5-m Isaac Newton Telescope (INT) in Spain over the wavelength range 3525-7500 $\angstrom$ at a resolution of 2.5 $\angstrom$, which is well suited to the wavelength range covered by the OzDES spectra. The full approach is described by \cite{2017_Cappellari}.

We run the stacked spectra through pPXF to extract information about the stellar population. Figure \ref{fig:ppxf} shows the resulting high quality fit obtained using pPXF on an example OzDES stack. The output consists of the relative flux of the input spectrum, the best fit from the stellar templates, gas emission lines, and fit residuals. We also obtain weighted ages, metallicities and mean mass-to-light ratios of the best fit stellar population.
We note that an individual OzDES spectrum is not suited for using full spectrum fitting due to the low S/N of individual spectra. 

\section{Results} \label{section4}

\subsection{Trends between Hubble residual and host properties}

Here we explore trends between Hubble residuals and the properties of the hosts derived from DES photometry and OzDES spectroscopy. 
We apply a redshift cut ($z \leq 0.8$), and then split the sample into 12 bins between $-0.50 < \Delta \mu < 0.50$, with a similar number of objects per bin. The bin size is chosen so that the S/N of the stacked spectra is sufficient for a reliable pPXF fit (Figure \ref{fig:stacks}).

pPXF fits for both the continuum that comes from stars and nebula emission lines from the ISM. As the spectra are normalised by the continuum before stacking, the line fluxes cannot be transformed into star formation rates. Instead, the line fluxes are more closely related to the specific star formation rate (sSFR); see Section \ref{Coadding} for a discussion on how the OzDES data are scaled before coadding and stacking. 


In Figure \ref{fig:finer_hr_binning}, we plot each of the host galaxy properties obtained from pPXF fits to the stacked spectra against Hubble residuals. We also show the average stellar mass for each stack obtained from the photometry. The outer errorbars for each point were obtained from bootstrap resampling, with the $1\sigma$ confidence intervals reported in 
Table \ref{table:ppxf_output_table}. Given that these errors are small ($\lesssim 0.01$ mag), they can be safely ignored. We fit a linear relation to each property instead of a step function to better capture the trends in our data, the results of which are summarized in Table \ref{table:ppxf_output}.

\begin{table*}
\centering
\caption{For each of the galaxy properties we show the slope of the best straight line fit, significance and the Pearson correlation coefficient, to give an indication of the strength of any trend. The most significant trend is between stellar mass and Hubble residual. The trend between the strength of the [OII] line and Hubble residual is the next most significant. The remaining properties show no obvious trends.}
\begin{tabular}{ccccc}
\hline
\textbf{Host Property} & \textbf{Slope} &
\textbf{Significance ($\sigma$)} & \textbf{Correlation}
\\ 
\hline
log Stellar Age (yrs) & $-0.10 \pm 0.10$ & 0.99 & 0.03 \\
Mass-to-light ratio & $-0.25 \pm 0.35$ & 0.95 & 0.06 \\ 
Metallicity [M/H] & $-0.044 \pm 0.033$ & 1.33 & $-$0.23 \\ 
Equivalent Width [OII] ($\angstrom$) & $-8.96 \pm 2.02$ & 4.44 &  $-$0.68 \\
log Stellar Mass $\mathrm{M_{\bigodot}}$ & $-0.86 \pm 0.12$ & 7.2 & $-$0.88 \\
\hline
\end{tabular}
\label{table:ppxf_output}   
\end{table*}

\begin{figure*}
    \centering
    \includegraphics[scale=0.175]{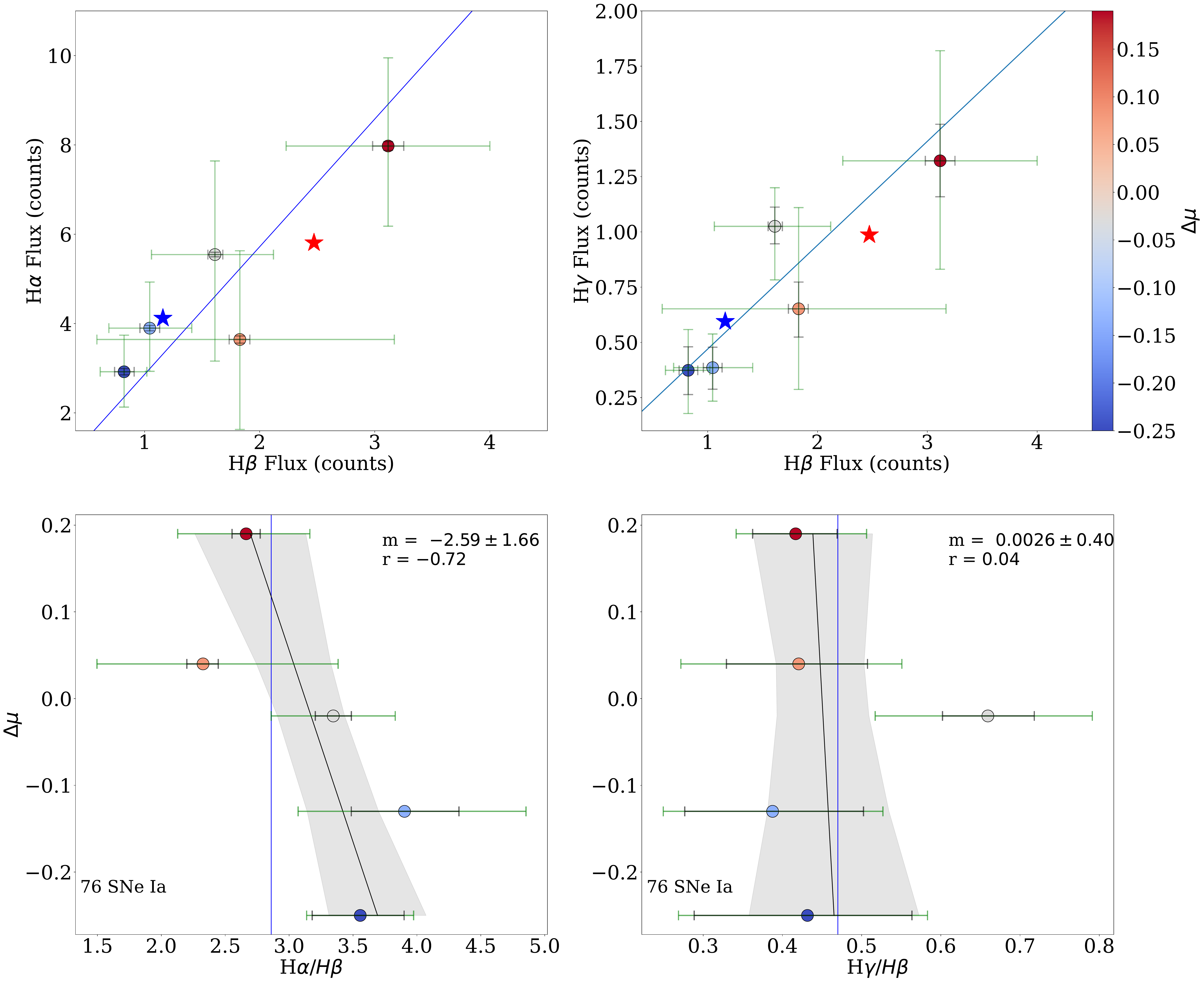}
    \caption{Top: The ratio of emission line strengths for the low redshift sample, $z \leq 0.35$, with the blue line corresponding to the theoretical Balmer ratio in each case. The Balmer lines tend to be stronger for the fainter SNe Ia. The red/blue stars show the mean line strengths for the positive/negative Hubble residuals. 
    Bottom: Balmer ratio against Hubble residual for the same sample of SN Ia hosts. We see a possible trend between Hubble residual and H$\alpha$/H$\beta$ with a slope of $-2.59\pm1.66$ and $r = -0.72^{-0.15}_{+0.22}$ and no trend for H$\gamma$/H$\beta$ with a slope of $-0.00\pm0.40$, and $r = 0.04^{+0.37}_{-0.34}$. }
    \label{fig:balmer_bd_ab}
\end{figure*}

The host stellar mass trend is the most significant, although interestingly no trend is present in the mass-to-light ratio, suggesting the correlation may not be heavily influenced by the stellar populations themselves. Rather than a sharp transition at a single mass, we see a gradual trend.

From the OzDES galaxy spectra, [OII] shows the clearest trend, with the positive residuals residing in galaxies with larger [OII] line fluxes and therefore with higher sSFRs. As noted above, the spectra are normalised by the continuum before stacking, so line fluxes are more closely related to the specific star formation rate (sSFR) than the absolute SFR.

In Figure \ref{fig:EW}, we plot the EW of [OII] against the EW of $H\beta$, finding a clear correlation between the two lines. As both lines are sensitive to star formation, the EWs are a measure of the sSFR, and the correlation between the two EWs is expected. 
Note how the EWs of both lines  correlate with  Hubble residual. Hosts with large EWs and therefore larger specific star formation rates also tend to host SN Ia with more positive Hubble residuals.

The trend between Hubble residual and the sSFR has been seen in previous studies (\citealp{Sullivan_2010, 2021Galbany}) and found to be the strongest step when using local sSFR (\citealp{Rigault_2020,2021Briday}). Lastly we find no trend with stellar population age and only a weak trend with metallicity \citep{2016campbell}, where metal poor host galaxies contain more positive residuals \citep{Childress_2013}.

\begin{table}
\caption{Similar to Table \ref{table:ppxf_output_table}, containing the mean values and resulting errorbars from Figure \ref{fig:balmerdust}.}
\begin{tabular}{cccc}
\hline
$\Delta \mu$ & $H\gamma$ & $H\beta$ & Balmer Ratio ($H\gamma/H\beta)$ \\
\hline
$-0.37$ & $ 0.76 \pm 0.12 $ & $ 1.36 \pm 0.18 $ & $ 0.56 \pm 0.07 $ \\
$-0.23$ & $ 0.63 \pm 0.17 $ & $ 1.19 \pm 0.31 $ & $ 0.53 \pm 0.07 $ \\
$-0.17$ & $ 0.50 \pm 0.15 $ & $ 1.10 \pm 0.26 $ & $ 0.45 \pm 0.07 $ \\
$-0.12$ & $ 0.51 \pm 0.21 $ & $ 1.03 \pm 0.52 $ & $ 0.52 \pm 0.12 $ \\
$-0.06$ & $ 0.87 \pm 0.24 $ & $ 1.86 \pm 0.34 $ & $ 0.47 \pm 0.08 $ \\
$-0.01$ & $ 0.90 \pm 0.20 $ & $ 1.63 \pm 0.36 $ & $ 0.56 \pm 0.07 $ \\
0.02 & $ 0.60 \pm 0.16 $ & $ 1.38 \pm 0.45 $ & $ 0.45 \pm 0.07 $ \\
0.06 & $ 1.29 \pm 0.21 $ & $ 3.15 \pm 0.71 $ & $ 0.42 \pm 0.06 $ \\
0.10 & $ 1.45 \pm 0.17 $ & $ 2.68 \pm 0.28 $ & $ 0.54 \pm 0.04 $ \\
0.16 & $ 0.58 \pm 0.25 $ & $ 1.42 \pm 0.58 $ & $ 0.41 \pm 0.10 $ \\
0.23 & $ 0.67 \pm 0.27 $ & $ 1.76 \pm 0.64 $ & $ 0.37 \pm 0.07 $ \\
0.39 & $ 0.70 \pm 0.28 $ & $ 1.88 \pm 0.59 $ & $ 0.37 \pm 0.08 $ \\
\hline
\end{tabular}
\label{table:dust_output}   
\end{table}

\subsection{Dust}
To further investigate these trends, we also probe the effect that extrinsic factors such as dust may have on the luminosity of SNe Ia after standard corrections have been applied. 

A diagnostic measure of reddening due to dust is the Balmer decrement, representing the ratio between hydrogen nebular emission lines (H$\alpha$, H$\beta$ and H$\gamma$, etc) \citep{BalmerDecrement}. For Case B recombination H$\alpha$/H$\beta$ = 2.86 and H$\gamma$/H$\beta$ = 0.468 \citep{1989osterbrock}. Interstellar dust attenuation significantly impacts ultraviolet and visible wavelengths more than longer wavelengths. The amount of reddening can be determined by comparing the observed and theoretical Balmer decrements and any deviation will give an indication of the amount of reddening and, given an attenuation law, the amount of attenuation due to dust.

\subsubsection{The Balmer Decrement}
The most common Balmer line ratio used to probe  reddening due to dust is H$\alpha$/H$\beta$, where values above the theoretical ratio 2.86 indicate the degree of reddening.  However, we need to limit our selection to $z \leq 0.35$ (which we define as the low redshift sample) due to H$\alpha$ becoming redshifted out of the wavelength range of our spectra. This impacts the sample size of each stack, with a small number of objects to stack resulting in a lower S/N in comparison to the larger sample. Measuring the emission line for $H\gamma$, we can also make use of the second Balmer line ratio H$\gamma$/H$\beta$, where values below 0.468 give an indication of the degree of reddening.

\begin{figure} 
    \centering
    \includegraphics[scale=0.175]{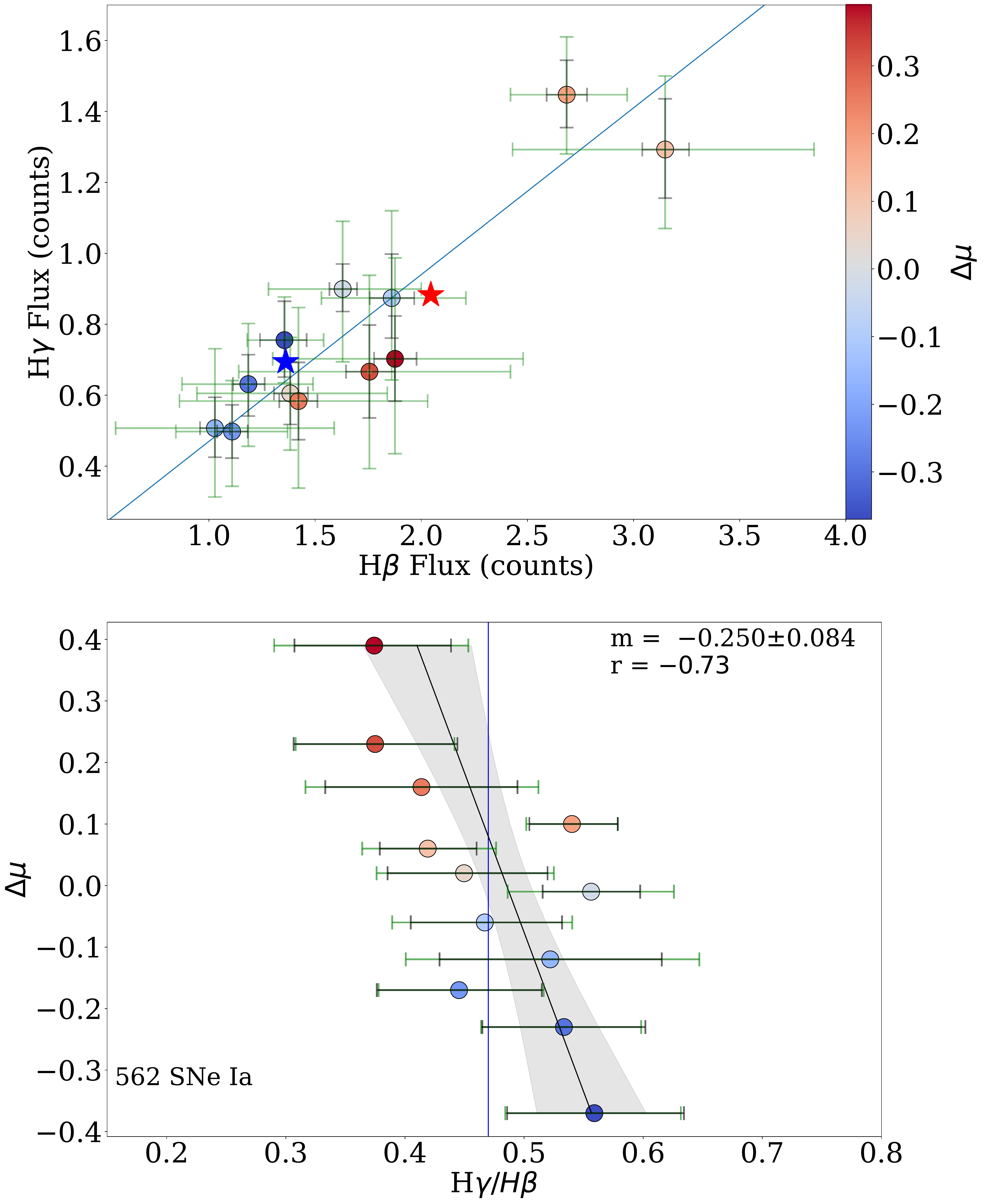}
    \caption{Top: The ratio of emission lines (H$\beta$ and H$\gamma$) for the sample described in Section \ref{section4}, where $z \leq 0.80$. The blue line represents the expected Balmer decrement for Case B recombination of 0.468 for H$\gamma$/H$\beta$. The mean line strengths for positive and negative Hubble residuals are shown as red/blue stars respectively. Below: The average Hubble residual against Balmer ratio. Similarly, the blue line corresponds to the expected value, and we observe that the hosts with more positive (fainter) Hubble residual have lower H$\gamma$/H$\beta$ ratios, indicating more reddening in these stacks. Fitting the data we obtain a slope of $-0.176\pm0.074$, where $r = -0.52^{-0.12}_{+0.14}$.}
    \label{fig:balmerdust}
\end{figure}

In Figure~\ref{fig:balmer_bd_ab}, we plot the strength of the H$\alpha$ line against the strength of H$\beta$, and make a similar plot for H$\gamma$ and H$\beta$. In both plots, fainter SNe Ia tend to appear in galaxies with larger line strengths. This is similar to the trend seen in Figure \ref{fig:EW}. Galaxies with larger line strengths, and therefore high specific star formation rates\footnote{When processing the OzDES data, the spectra are normalised to 1 over the observer-frame wavelength range 5800\AA\, to 8600\AA}. Hence the line strengths are a measure of the sSFR and not the SFR, are collectively dimmer.


In the lower panels, we plot the Hubble residual versus line ratio. If reddening is not present, then all points will lie on the vertical blue lines. We see a very weak trend with low significance between Hubble residual and the ratio of H$\alpha$ and H$\beta$, and no trend for 
H$\gamma$ and H$\beta$. The weak trend seen using H$\alpha$ and H$\beta$ is opposite to what one expects, if dust affects both the SN Ia luminosities and the Balmer line ratio.

The H$\gamma$ and H$\beta$ line ratio provides access to a larger sample. The increased sample size gives a more significant result. In Figure \ref{fig:balmerdust}, the averaged trend across the stacks is that the fainter SNe Ia tend to have smaller ratios, which is consistent with the notion of attenuation and reddening by dust. 

We also see the same trend between residual and line strength. Sources with more positive residuals have stronger lines and this trend is evident for all three lines. This is consistent with the trend between the strength of [OII] and Hubble residual shown in Figure \ref{fig:finer_hr_binning}, as the strengths of all four lines are related to the specific star formation rate.

We note that our observed Balmer ratio correlation with Hubble residual does not offer a reddening correction for each individual object.

\begin{figure}
    \centering
    \includegraphics[scale=0.175]{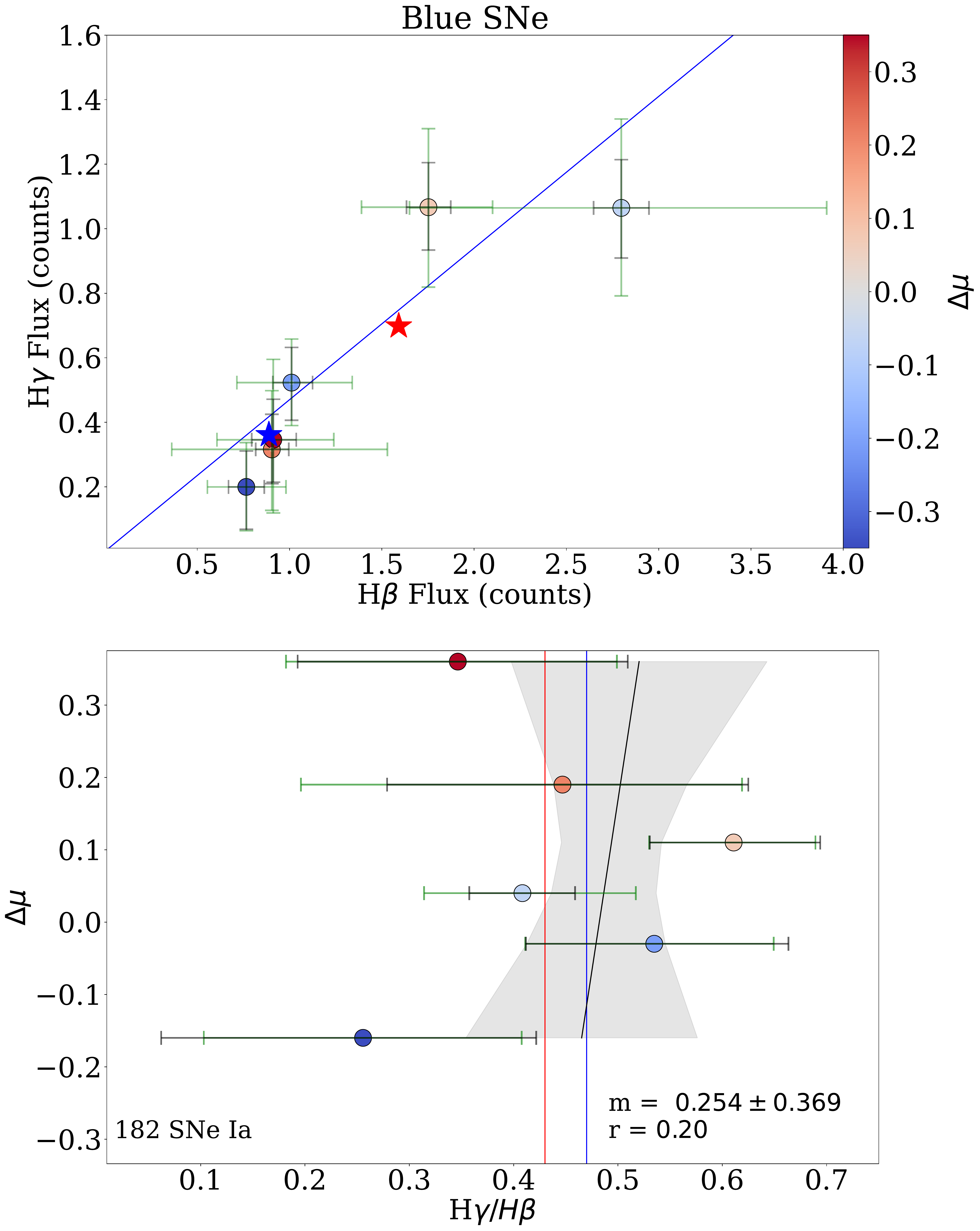}
    \caption{Host galaxy sample for blue SNe Ia ($c < -0.025$) and then binned by Hubble residual. Fitting the data we obtain a slope of $0.254\pm0.369$, where $r = 0.20^{+0.30}_{-0.33}$. The blue line represents the theoretical Balmer ratio (\citealp{1989osterbrock}), while the average Balmer ratio, shown as the red line, is $0.43\pm0.052$. Similar to previous figures, mean line strengths for positive/negative Hubble residuals are shown as red/blue stars in the top panel.}
    \label{fig:balmerblue}
\end{figure}

\begin{figure}
    \centering
    \includegraphics[scale=0.175]{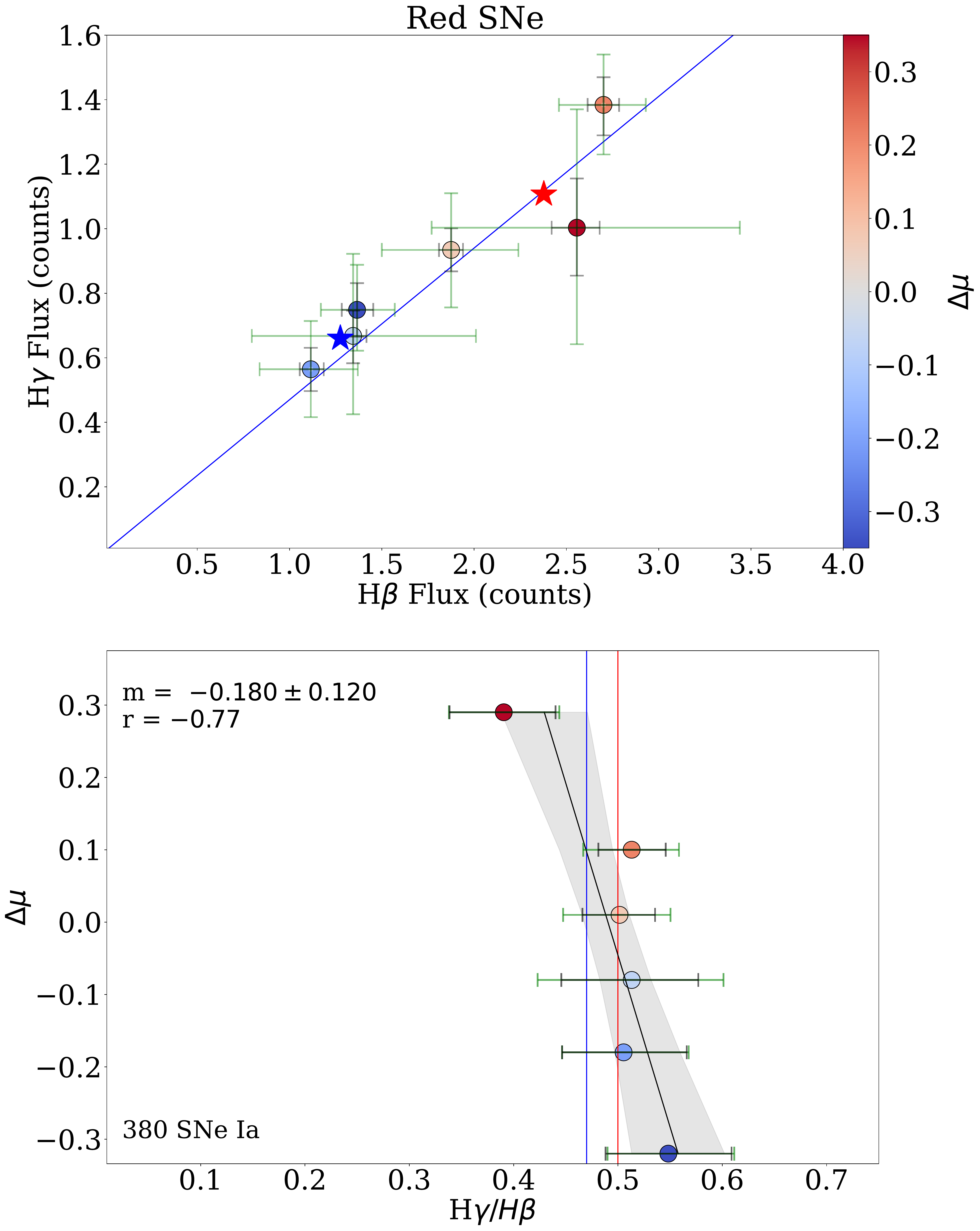}
    \caption{Host galaxy sample split into redder SNe Ia ($c > -0.025$) and then binned by Hubble residual. Fitting the data we obtain a slope $-0.180\pm0.11$, where $r = -0.77^{-0.11}_{+0.19}$. The blue line represents the theoretical Balmer ratio (\citealp{1989osterbrock}), while the average Balmer ratio, shown as the red line, is $0.50\pm0.022$. Similar to previous figures, mean line strengths for positive/negative Hubble residual are shown as red/blue stars in the top panel.}
    \label{fig:balmerred}
\end{figure}

\subsubsection{SN Colour and Reddening}
We next split our sample by SN Ia colour and then bin by Hubble residual to examine if the trends noted in the previous section depend on SN Ia colour. A cut is made at $c = -0.025$, using the same dividing line as \cite{brout2021dust}. The results are shown in Figures~\ref{fig:balmerblue} and \ref{fig:balmerred}.

There are two noteworthy trends. Firstly, the redder SNe Ia tend to reside in galaxies that have stronger line strengths compared with the bluer SNe Ia, indicating higher specific star formation rates in galaxies that host red SNe Ia. The mean line fluxes of SN Ia hosts for the blue/red samples are (0.59, 0.88) for $H\gamma$ and (1.36, 1.83) for $H\beta$. 

Secondly, the trend between line ratios and Hubble residuals disappears for the blue SNe Ia and remains, albeit very weakly for red SNe Ia.


\section{Discussion}
\label{section5}

\subsection{Trends with emission line strengths}

We find that there are trends between Hubble residual and the strengths of [OII], $H\gamma$, $H\beta$ and $H\alpha$ emission lines of the host galaxies. We also find  trends between the EWs of  [OII] and $H\beta$ with Hubble residual. SNe Ia are fainter in host galaxies that have stronger emission lines and larger EWs after the standard corrections for colour and light curve width have been applied to the SNe Ia.

The EWs of the [OII] and $H\beta$ emission lines depend on the specific star formation rate (sSFR). Although other factors, e.g. metallicity, can affect the EW of [OII], generally speaking, galaxies with larger EWs have higher sSFRs and are more actively forming stars. The line strengths of [OII] and the Balmer emission lines in the OzDES spectra also depend on the sSFR, as the OzDES spectra are normalized by the continuum before coadding and stacking.

The trend between Hubble residual and the sSFR of the host galaxy has been noted by a number of other studies \citep{Sullivan_2010,Lampeitl_2010, 2011d'andrea, Childress_2013, Rigault_2020}. It has been detected using broad band photometry and spectroscopy.

Using broad band photometry of SNe Ia from the Supernova Legacy Survey, \citet{Sullivan_2010} find that SNe Ia in galaxies with low sSFRs are  brighter after light curve correction than galaxies with high sSFRs. When modelled as a step at $\mathrm{log (sSFR)}=-9.7$, the step is $\sim 0.05$ mag. Using spectroscopy of SNe Ia from the Sloan Digital Sky Survey, \citet{2011d'andrea} also find that SNe Ia are brighter in galaxies that have low sSFRs.

Due to the nature of the observations and the redshifts of the hosts, these earlier works were not sensitive to the properties of the hosts in the region immediately surrounding the SN. This also applies to the OzDES data analysed in this paper. Most OzDES galaxies subtend an angular size of an arc-second, which is smaller than the 2 arc-second diameter of the 2dF fibres. Hence OzDES spectra are not sensitive to the properties of the hosts in the region immediately surrounding the SN. Instead, OzDES and these earlier works measure global properties of the hosts.

\citet{Rigault_2020}, using IFU spectroscopy of the hosts of nearby SNe Ia, found a trend between Hubble residuals and the properties of host galaxies in a projected 1~kpc region surrounding the site of SNe. When modelled as a step function with the step located at $\mathrm{log (LsSFR)}=-9.7$, where LsSFR is the local sSFR, they found a step of $0.163\pm 0.29$ mag. This is larger than the step that is measured using host galaxy mass and is the largest step measured to date.

We do not see evidence for a trend between Hubble residuals and galaxy age. As younger galaxies tend to have higher sSFRs, the lack of a trend is surprising.

As one goes to higher redshifts, the global sSFR increases, leading to a potential evolution in the SN population. At higher redshifts, there will be a higher fraction of SNe Ia in galaxies that have sSFRs above the step. This naturally leads to the question of how this may bias the estimation of cosmological parameters, such as the Hubble constant and the dark energy equation of state parameter, a question that others have examined (\citealp{Childress_2013, Rigault_2020, 2022Brout}) and a topic that we will come back to in section \ref{ImpactonCosmo}.


\subsection{Evidence for dust}

We find no evidence for a trend between the properties of the host measured from the continuum of the spectra and Hubble residuals. This includes age, which is surprising, as line strengths and EWs are a measure of the sSFR which is an indicator of age. Galaxies with high sSFR are younger than galaxies with low sSFR. Could the trend be driven by another property related to the sSFR?

The amount of dust and the properties of the dust also correlate with host galaxy properties. More massive galaxies tend to be more attenuated by dust and have shallower attenuation laws \citep{2018Salim}. Dust properties also correlate with sSFR \citep{2017Orellana,2021Triani}. Galaxies with higher sSFR have higher dust-to-stellar mass ratios.


We used the Balmer line ratios in the stacked spectra as a measure of the reddening by dust and searched for correlations between the Balmer line ratios and Hubble residuals. There is some evidence that the H$\gamma$/H$\beta$ line ratio correlates with Hubble residual, but the trend is not very significant. Fainter SNe Ia tend to lie in galaxies that have smaller H$\gamma$/H$\beta$ line ratios, consistent with the idea that the galaxies that hosted these SNe Ia are more affected by dust. The trend appears to be driven by the redder SNe Ia. When the sample is split by SN Ia colour, the evidence for a trend persists for the red SNe Ia (defined here as those with $c\geq -0.025$), but is absent for the blue SNe Ia.

As noted at the beginning of this section, we also found a clear trend between the strength of the emission lines and Hubble residuals. When we split the sample according to SN Ia color, we find that red SNe Ia - defined again as those with $c\geq -0.025$ - occur in galaxies with stronger emission lines.
These galaxies have higher sSFRs, which have been shown to correlate with dust-to-stellar mass ratios  \citep{2017Orellana,2021Triani}. Similar trends have been found between global/local (U - R) colour and DES5YR Hubble residuals \citep{kelsey2022}.

Overall, there are hints that dust is playing a role, but the trends are not very significant. Yet, dust is a key part of the BS21 model.

In the BS21 model, the amount of attenuation and the attenuation law, as parameterised by $R_V$, depends on host galaxy mass. More massive galaxies have shallower attenuation laws but more dust. The model successfully explains the relationships between Hubble scatter and  Hubble residuals with SN Ia colour and explains the mass step as a result of the correlation between the galaxy mass and the properties of the dust. The model is consistent with other observations \cite{2018Salim}.

The lack of a very significant trend between the amount of reddening inferred from the spectra and Hubble residual is in part due to the limitations of the approach. The reddening inferred from the spectra, which applies to the whole galaxy, may only be weakly correlated to the amount of reddening affecting the supernova.

\subsection{Limitations of this analysis and future work}

While a number of trends have been shown to exist on stacked spectra, it is unclear how strong they are on individual spectra. This is a limitation of our work. Having higher S/N spectra would allow full spectrum fitting of individual hosts. This is only possible for a small fraction of the host galaxies observed by OzDES and the numbers are too small for clear trends to emerge.

Future surveys, such as TIDES \citep{2019Swann} on the VISTA telescope at the Paranal Observatory in Chile will obtain spectra of tens of thousands of galaxies that host SNe Ia discovered by the Legacy Survey of Space and Time. This is two orders of magnitude more hosts than analysed here, and the number will be sufficient to see if the trends seen in the stacked spectra are also present when the hosts are analysed individually.

Furthermore, as the seeing is at least a factor of two better than the seeing at Siding Spring, where the AAT is located, and the fibre diamater smaller, the data will probe smaller regions of the host galaxy. For nearby galaxies, it will be possible to measure local properties of the host, as done in \citet{2018Galbany} and \citet{Rigault_2020}.

Another limitation of our work is the weighting we apply when stacking spectra. More weight is given to brighter objects. We repeated the analysis without weighting and found that the trends remain although with reduced significance.

We also note the insensitivity of the results when $\alpha$ and $\beta$ are allowed to vary within their uncertainties.


\subsection{Impact on Cosmology}\label{ImpactonCosmo}

\subsubsection{The impact on dark energy equation of state}

SNe Ia constrain the dark energy equation of state parameter, $w$, by comparing the luminosity of nearby SNe Ia to distant SNe Ia. In the Pantheon+ SN sample \citep{2022Brout}, distant SNe Ia extend up to $z\sim 2$, which corresponds to a lookback time of $\sim 10$~Gyr. Over that time, galaxies have evolved significantly. Compared to galaxies today, galaxies 10 Gyr ago were much less massive, less metal rich, bluer, and much more actively forming stars. All of these properties have been shown to correlate with SN Ia Hubble residuals after the standard corrections for light curve width and colour have been applied.

Furthermore, the mean dust attenuation law and the amount of attenuation from dust are also expected to change with increasing look-back time. In high-redshift galaxy analogs (galaxies that sit above the star formation mains sequence), the dust attenuation law is steeper that the law in the Milky Way \citep{2018Salim}. In this paper, we found a marginal trend between Hubble residuals and the Balmer decrement, an indicator of reddening by dust.

While it is not yet clear what drives these trends, left uncorrected, these trends will result in a bias in $w$. For example, if the age of the progenitor is driving these trends then \citet{Childress_2013} find that distant SNe Ia at $z \sim 1$ could be as much as 0.04 mag fainter after light curve and colour correction, corresponding to a change $w$ that is double the current statistical uncertainty in $w$.

A change in $w$ also results from changes in how the scatter in SNe Ia luminosities after standardisation are modelled. The scatter results in some SNe Ia being preferentially selected over others, which results in a bias that depends, to first order, on redshift. The magnitude of the bias is modelled through detailed simulations \citep{2016Scolnic} and is applied to the distance modulus (see Eq.~2). Over the past decade, the  scatter has been modelled phenomenologically with achromatic and chromatic terms contributing to the scatter. The two most common models are the G10 model \citep{G10}, where 80\% of the scatter is achromatic and the C11 model \citep{C11} where 20\% of the scatter is achromatic.

BS21 provide a more physically motivated model for the scatter. The BS21 model allows for an intrinsic relation between SN Ia colour and luminosity and the effects of dust in the host galaxy. In this model, the dust attenuation law, as described by $R_V$, and the amount of attenuation are allowed to vary and are described by distributions that depend on the properties of the hosts. BS21 find that the mass step is a natural result of the different properties of dust in low and high mass galaxies. BS21 find biases as large as 4\% in $w$ between their treatment of the scatter and the treatment of the scatter in the G10 and C11 models. The difference is similar to the statistical uncertainty in $w$ \citep{2022Brout}.

\subsubsection{The impact on $H_0$}

For $H_0$, the absolute brightness of SNe Ia is required. The brightness is calibrated in galaxies that have hosted SNe Ia and are near enough for the distances to be measured independently from the SNe Ia using, for example, Cepheids \citep{2022Riess}, or the tip of the red-giant branch (TRGB) \citep{freedman2021}. The SNe Ia that are used to measure the $H_{0}$ differ from the ones that are used to calibrate the absolute SN Ia brightness. As the SNe Ia between the two samples differ, so too will the hosts. There is therefore the potential for host-dependent luminosity variations to bias $H_{0}$.

Additional $H_{0}$ calibrator galaxies will provide improvements in the zero-point calibration of SNe Ia used in determining $H_{0}$. However, the number of future candidates will be limited due to the low rate of SN Ia explosions we can detect in nearby galaxies which also contain Cepheids, currently at a rate of $\sim$1/year. Recently, an increased sample size of 42 $H_{0}$ calibrator galaxies was utilised in the Cepheid approach of calibrating SNe Ia and this reduced each individual error contribution to below $1\%$ \citep{2022Riess}. Cepheids are known to reside in late-type galaxies and this could introduce a bias in the calibration of SNe Ia when comparing to a range of galaxy types at higher redshift which do not contain Cepheids. However, \cite{2016Riess} found that by using a homogeneous sample of late-type galaxies into the Hubble flow, there was no significant change to the uncertainty in measuring $H_{0}$. This is due to the much larger Hubble flow sample in comparison to the calibrator galaxies. Recent work on SN Ia sample selection also suggests the benefit of a SNe Ia sample containing bluer SNe Ia as they are less affected by dust and may be more suited for measuring cosmological parameters (\citealp{2021Kelsey, gonzalex2020, meldorf2022}).

As galaxies evolve with redshift, this means the stellar populations of galaxies into the Hubble flow will differ from nearby $H_{0}$ calibrator galaxies. As discussed earlier, the sSFR is the most significant Hubble residual step found to date after standard light curve corrections have been applied \citep{Rigault_2020}. Recently, \cite{2022Brout} found a sSFR step of $0.031\pm0.011$ in the Pantheon+ SN sample, which then reduced to $0.008\pm0.011$ after applying dust and mass bias corrections (BS21). This suggests that the link between host galaxy mass and dust properties accounts for correlations with host galaxy properties such as the sSFR.

\cite{2022Riess} suggest that while SNe Ia derived distances will be improved by uncovering host galaxy correlations after standardisation of SNe Ia, increasing the size and cross calibrating the calibrator and Hubble flows samples will help mitigate these potential uncertainties in measuring $H_{0}$. 


\subsubsection{Modifying the Tripp equation}

Modifying the Tripp equation (Eq.~2) to include the effects of dust explicitly for each SN Ia may be difficult to do in practice with current SN Ia samples. In addition to the determining the SN Ia colour law and the coefficient that relates the strength of correlation between intrinsic colour and luminosity, one needs to determine the dust attenuation law, as parameterised by $R_V$ or some other parameterisation, and the amount of attenuation. Most SNe Ia are observed over a limited wavelength range and in a handful of broadband filters, so it may not be possible to provide meaningful constraints on all the parameters.  Most SNe Ia lack observations in the rest frame near-IR, where the impact of dust is smallest and where it is possible, once combined with data in the rest-frame optical, to provide meaningful constraints \citep{2021johansson}. 

A less challenging approach may be to use the host mass to set $R_V$ and the amount of attenuation to their median values both in the training of the SALT2 light curve model and the fitting of the SN Ia light curves. However, detailed simulations will still be needed to correct for biases in the Hubble diagram.

\section{Conclusion}

Multiple studies have demonstrated that the scatter in the SN Ia Hubble diagram can be reduced by incorporating properties of the host galaxy, such as host galaxy mass, sSFR, and metallicity. Most of these studies have used broad band photometry to infer host properties.

Using a sample 625 SN Ia host galaxies from the DES 5-year photometrically confirmed sample, we search for trends between Hubble residuals and the properties of host galaxies inferred from stacked spectra. We see a clear trend between the strength of the emissions lines (an indicator of the sSFR) and Hubble residual, but find no significant trends between properties inferred from the stellar continuum, such as metallicty, mass-to-light ratio, and age.

We further examine the stacked spectra, searching for a trend between the Balmer decrement (an indicator of reddening by dust) and Hubble residual. We find a marginally significant trend between the H$\gamma$/H$\beta$ line ratio and Hubble residual. The trend is only present in the redder SNe Ia. This is consistent with the notion that redder SNe Ia are more affected by dust. However, the trends are marginal and a larger sample of SN Ia hosts is required to confirm them.

In the near future, 4MOST will obtain the spectra of tens of thousands of SN Ia host galaxies and will be able to examine the trends detected here with much greater statistical significance.

\section*{Acknowledgements}
The author Mitchell Dixon would like to acknowledge support through an Australian Government Research Training Program Scholarship. This research was supported by the Australian Research Council The Centre of Excellence for Dark Matter Particle Physics (CDM; project number CE200100008) and the Australian Research Council Centre of Excellence for Gravitational Wave Discovery (OzGrav; project number CE170100004). This project/ publication was made possible through the support of a grant from the John Templeton Foundation. The authors gratefully acknowledge this grant ID 61807, Two Standard Models Meet. The opinions expressed in this publication are those of the author(s) and do not necessarily reflect the views of the John Templeton Foundation.

Funding for the DES Projects has been provided by the U.S. Department of Energy, the U.S. National Science Foundation, the Ministry of Science and Education of Spain, 
the Science and Technology Facilities Council of the United Kingdom, the Higher Education Funding Council for England, the National Center for Supercomputing 
Applications at the University of Illinois at Urbana-Champaign, the Kavli Institute of Cosmological Physics at the University of Chicago, 
the Center for Cosmology and Astro-Particle Physics at the Ohio State University,
the Mitchell Institute for Fundamental Physics and Astronomy at Texas A\&M University, Financiadora de Estudos e Projetos, 
Funda{\c c}{\~a}o Carlos Chagas Filho de Amparo {\`a} Pesquisa do Estado do Rio de Janeiro, Conselho Nacional de Desenvolvimento Cient{\'i}fico e Tecnol{\'o}gico and 
the Minist{\'e}rio da Ci{\^e}ncia, Tecnologia e Inova{\c c}{\~a}o, the Deutsche Forschungsgemeinschaft and the Collaborating Institutions in the Dark Energy Survey. 

The Collaborating Institutions are Argonne National Laboratory, the University of California at Santa Cruz, the University of Cambridge, Centro de Investigaciones Energ{\'e}ticas, 
Medioambientales y Tecnol{\'o}gicas-Madrid, the University of Chicago, University College London, the DES-Brazil Consortium, the University of Edinburgh, 
the Eidgen{\"o}ssische Technische Hochschule (ETH) Z{\"u}rich, 
Fermi National Accelerator Laboratory, the University of Illinois at Urbana-Champaign, the Institut de Ci{\`e}ncies de l'Espai (IEEC/CSIC), 
the Institut de F{\'i}sica d'Altes Energies, Lawrence Berkeley National Laboratory, the Ludwig-Maximilians Universit{\"a}t M{\"u}nchen and the associated Excellence Cluster Universe, 
the University of Michigan, NSF's NOIRLab, the University of Nottingham, The Ohio State University, the University of Pennsylvania, the University of Portsmouth, 
SLAC National Accelerator Laboratory, Stanford University, the University of Sussex, Texas A\&M University, and the OzDES Membership Consortium.

Based in part on observations at Cerro Tololo Inter-American Observatory at NSF's NOIRLab (NOIRLab Prop. ID 2012B-0001; PI: J. Frieman), which is managed by the Association of Universities for Research in Astronomy (AURA) under a cooperative agreement with the National Science Foundation.

The DES data management system is supported by the National Science Foundation under Grant Numbers AST-1138766 and AST-1536171.
The DES participants from Spanish institutions are partially supported by MICINN under grants ESP2017-89838, PGC2018-094773, PGC2018-102021, SEV-2016-0588, SEV-2016-0597, and MDM-2015-0509, some of which include ERDF funds from the European Union. IFAE is partially funded by the CERCA program of the Generalitat de Catalunya.
Research leading to these results has received funding from the European Research
Council under the European Union's Seventh Framework Program (FP7/2007-2013) including ERC grant agreements 240672, 291329, and 306478.
We  acknowledge support from the Brazilian Instituto Nacional de Ci\^encia
e Tecnologia (INCT) do e-Universo (CNPq grant 465376/2014-2).

This manuscript has been authored by Fermi Research Alliance, LLC under Contract No. DE-AC02-07CH11359 with the U.S. Department of Energy, Office of Science, Office of High Energy Physics.

Based on data acquired at the Anglo-Australian Telescope, under program A/2013B/012. We acknowledge the traditional custodians of the land on which the AAT stands, the Gamilaraay people, and pay our respects to elders past and present.


\section*{Data Availability}
The DES-SN photometric SN Ia catalogue will be made available as part of the DES5YR SN cosmology analysis at https://des.ncsa.illinois.edu/releases/sn.
The OzDES-DR2 spectra utilised in this paper can be publically accessed at: https://docs.datacentral.org.au/ozdes/overview/dr2.


\typeout{}
\bibliographystyle{mnras}
\bibliography{ref} 

\begin{thebibliography}{}
\makeatletter
\relax
\def\mn@urlcharsother{\let\do\@makeother \do\$\do\&\do\#\do\^\do\_\do\%\do\~}
\def\mn@doi{\begingroup\mn@urlcharsother \@ifnextchar [ {\mn@doi@}
  {\mn@doi@[]}}
\def\mn@doi@[#1]#2{\def\@tempa{#1}\ifx\@tempa\@empty \href
  {http://dx.doi.org/#2} {doi:#2}\else \href {http://dx.doi.org/#2} {#1}\fi
  \endgroup}
\def\mn@eprint#1#2{\mn@eprint@#1:#2::\@nil}
\def\mn@eprint@arXiv#1{\href {http://arxiv.org/abs/#1} {{\tt arXiv:#1}}}
\def\mn@eprint@dblp#1{\href {http://dblp.uni-trier.de/rec/bibtex/#1.xml}
  {dblp:#1}}
\def\mn@eprint@#1:#2:#3:#4\@nil{\def\@tempa {#1}\def\@tempb {#2}\def\@tempc
  {#3}\ifx \@tempc \@empty \let \@tempc \@tempb \let \@tempb \@tempa \fi \ifx
  \@tempb \@empty \def\@tempb {arXiv}\fi \@ifundefined
  {mn@eprint@\@tempb}{\@tempb:\@tempc}{\expandafter \expandafter \csname
  mn@eprint@\@tempb\endcsname \expandafter{\@tempc}}}

\bibitem[\protect\citeauthoryear{{Betoule} et~al.,}{{Betoule}
  et~al.}{2014}]{Betoule2014}
{Betoule} M.,  et~al., 2014, \mn@doi [\aap] {10.1051/0004-6361/201423413},
  \href {http://adsabs.harvard.edu/abs/2014A%26A...568A..22B} {568, A22}

\bibitem[\protect\citeauthoryear{{Boone} et~al.,}{{Boone}
  et~al.}{2021}]{Boone2021b}
{Boone} K.,  et~al., 2021, \mn@doi [\apj] {10.3847/1538-4357/abec3b}, \href
  {https://ui.adsabs.harvard.edu/abs/2021ApJ...912...71B} {912, 71}

\bibitem[\protect\citeauthoryear{{Briday} et~al.,}{{Briday}
  et~al.}{2022}]{2021Briday}
{Briday} M.,  et~al., 2022, \mn@doi [\aap] {10.1051/0004-6361/202141160}, \href
  {https://ui.adsabs.harvard.edu/abs/2022A&A...657A..22B} {657, A22}

\bibitem[\protect\citeauthoryear{{Brout} \& {Scolnic}}{{Brout} \&
  {Scolnic}}{2021}]{brout2021dust}
{Brout} D.,  {Scolnic} D.,  2021, \mn@doi [\apj] {10.3847/1538-4357/abd69b},
  \href {https://ui.adsabs.harvard.edu/abs/2021ApJ...909...26B} {909, 26}

\bibitem[\protect\citeauthoryear{{Brout} et~al.,}{{Brout}
  et~al.}{2019a}]{2019BroutSMP}
{Brout} D.,  et~al., 2019a, \mn@doi [\apj] {10.3847/1538-4357/ab06c1}, \href
  {https://ui.adsabs.harvard.edu/abs/2019ApJ...874..106B} {874, 106}

\bibitem[\protect\citeauthoryear{{Brout} et~al.,}{{Brout}
  et~al.}{2019b}]{Brout_2019}
{Brout} D.,  et~al., 2019b, \mn@doi [\apj] {10.3847/1538-4357/ab08a0}, \href
  {https://ui.adsabs.harvard.edu/abs/2019ApJ...874..150B} {874, 150}

\bibitem[\protect\citeauthoryear{{Brout} et~al.,}{{Brout}
  et~al.}{2022}]{2022Brout}
{Brout} D.,  et~al., 2022, arXiv e-prints, \href
  {https://ui.adsabs.harvard.edu/abs/2022arXiv220204077B} {p. arXiv:2202.04077}

\bibitem[\protect\citeauthoryear{{Campbell}, {Fraser}  \& {Gilmore}}{{Campbell}
  et~al.}{2016}]{2016campbell}
{Campbell} H.,  {Fraser} M.,   {Gilmore} G.,  2016, \mn@doi [\mnras]
  {10.1093/mnras/stw115}, \href
  {https://ui.adsabs.harvard.edu/abs/2016MNRAS.457.3470C} {457, 3470}

\bibitem[\protect\citeauthoryear{{Cappellari}}{{Cappellari}}{2017}]{2017_Cappellari}
{Cappellari} M.,  2017, \mn@doi [\mnras] {10.1093/mnras/stw3020}, \href
  {https://ui.adsabs.harvard.edu/abs/2017MNRAS.466..798C} {466, 798}

\bibitem[\protect\citeauthoryear{{Cappellari} \& {Emsellem}}{{Cappellari} \&
  {Emsellem}}{2004}]{2004_Cappellari}
{Cappellari} M.,  {Emsellem} E.,  2004, \mn@doi [\pasp] {10.1086/381875}, \href
  {https://ui.adsabs.harvard.edu/abs/2004PASP..116..138C} {116, 138}

\bibitem[\protect\citeauthoryear{{Childress} et~al.,}{{Childress}
  et~al.}{2013}]{Childress_2013}
{Childress} M.,  et~al., 2013, \mn@doi [\apj] {10.1088/0004-637X/770/2/108},
  \href {https://ui.adsabs.harvard.edu/abs/2013ApJ...770..108C} {770, 108}

\bibitem[\protect\citeauthoryear{{Childress} et~al.,}{{Childress}
  et~al.}{2017}]{Childress_2017}
{Childress} M.~J.,  et~al., 2017, \mn@doi [\mnras] {10.1093/mnras/stx1872},
  \href {https://ui.adsabs.harvard.edu/abs/2017MNRAS.472..273C} {472, 273}

\bibitem[\protect\citeauthoryear{{Chotard} et~al.,}{{Chotard}
  et~al.}{2011}]{C11}
{Chotard} N.,  et~al., 2011, \mn@doi [\aap] {10.1051/0004-6361/201116723},
  \href {https://ui.adsabs.harvard.edu/abs/2011A&A...529L...4C} {529, L4}

\bibitem[\protect\citeauthoryear{{D'Andrea} et~al.,}{{D'Andrea}
  et~al.}{2011}]{2011d'andrea}
{D'Andrea} C.~B.,  et~al., 2011, \mn@doi [\apj] {10.1088/0004-637X/743/2/172},
  \href {https://ui.adsabs.harvard.edu/abs/2011ApJ...743..172D} {743, 172}

\bibitem[\protect\citeauthoryear{{DES Collaboration} et~al.,}{{DES
  Collaboration} et~al.}{2016}]{Abbott2016}
{DES Collaboration} et~al., 2016, \mn@doi [\mnras] {10.1093/mnras/stw641},
  \href {https://ui.adsabs.harvard.edu/abs/2016MNRAS.460.1270D} {460, 1270}

\bibitem[\protect\citeauthoryear{{Dhawan}, {Brout}, {Scolnic}, {Goobar},
  {Riess}  \& {Miranda}}{{Dhawan} et~al.}{2020}]{Dhawan_2020}
{Dhawan} S.,  {Brout} D.,  {Scolnic} D.,  {Goobar} A.,  {Riess} A.~G.,
  {Miranda} V.,  2020, \mn@doi [\apj] {10.3847/1538-4357/ab7fb0}, \href
  {https://ui.adsabs.harvard.edu/abs/2020ApJ...894...54D} {894, 54}

\bibitem[\protect\citeauthoryear{{Dom{\'\i}nguez} et~al.,}{{Dom{\'\i}nguez}
  et~al.}{2013}]{BalmerDecrement}
{Dom{\'\i}nguez} A.,  et~al., 2013, \mn@doi [\apj]
  {10.1088/0004-637X/763/2/145}, \href
  {https://ui.adsabs.harvard.edu/abs/2013ApJ...763..145D} {763, 145}

\bibitem[\protect\citeauthoryear{{Flaugher} et~al.,}{{Flaugher}
  et~al.}{2015}]{Flaugher2015}
{Flaugher} B.,  et~al., 2015, \mn@doi [\aj] {10.1088/0004-6256/150/5/150},
  \href {https://ui.adsabs.harvard.edu/abs/2015AJ....150..150F} {150, 150}

\bibitem[\protect\citeauthoryear{{Freedman}}{{Freedman}}{2021}]{freedman2021}
{Freedman} W.~L.,  2021, \mn@doi [\apj] {10.3847/1538-4357/ac0e95}, \href
  {https://ui.adsabs.harvard.edu/abs/2021ApJ...919...16F} {919, 16}

\bibitem[\protect\citeauthoryear{{Galbany} et~al.,}{{Galbany}
  et~al.}{2018}]{2018Galbany}
{Galbany} L.,  et~al., 2018, \mn@doi [\apj] {10.3847/1538-4357/aaaf20}, \href
  {https://ui.adsabs.harvard.edu/abs/2018ApJ...855..107G} {855, 107}

\bibitem[\protect\citeauthoryear{{Galbany} et~al.,}{{Galbany}
  et~al.}{2022}]{2021Galbany}
{Galbany} L.,  et~al., 2022, \mn@doi [\aap] {10.1051/0004-6361/202141568},
  \href {https://ui.adsabs.harvard.edu/abs/2022A&A...659A..89G} {659, A89}

\bibitem[\protect\citeauthoryear{{Gonz{\'a}lez-Gait{\'a}n}, {de Jaeger},
  {Galbany}, {Mour{\~a}o}, {Paulino-Afonso}  \&
  {Filippenko}}{{Gonz{\'a}lez-Gait{\'a}n} et~al.}{2021}]{gonzalex2020}
{Gonz{\'a}lez-Gait{\'a}n} S.,  {de Jaeger} T.,  {Galbany} L.,  {Mour{\~a}o} A.,
   {Paulino-Afonso} A.,   {Filippenko} A.~V.,  2021, \mn@doi [\mnras]
  {10.1093/mnras/stab2802}, \href
  {https://ui.adsabs.harvard.edu/abs/2021MNRAS.508.4656G} {508, 4656}

\bibitem[\protect\citeauthoryear{{Gupta} et~al.,}{{Gupta}
  et~al.}{2016}]{Gupta_2016}
{Gupta} R.~R.,  et~al., 2016, \mn@doi [\aj] {10.3847/0004-6256/152/6/154},
  \href {https://ui.adsabs.harvard.edu/abs/2016AJ....152..154G} {152, 154}

\bibitem[\protect\citeauthoryear{{Guy} et~al.,}{{Guy} et~al.}{2010}]{G10}
{Guy} J.,  et~al., 2010, \mn@doi [\aap] {10.1051/0004-6361/201014468}, \href
  {https://ui.adsabs.harvard.edu/abs/2010A&A...523A...7G} {523, A7}

\bibitem[\protect\citeauthoryear{{Hartley} et~al.,}{{Hartley}
  et~al.}{2022}]{hartley2020dark}
{Hartley} W.~G.,  et~al., 2022, \mn@doi [\mnras] {10.1093/mnras/stab3055},
  \href {https://ui.adsabs.harvard.edu/abs/2022MNRAS.509.3547H} {509, 3547}

\bibitem[\protect\citeauthoryear{{Hoormann} et~al.,}{{Hoormann}
  et~al.}{2019}]{hoorman2019}
{Hoormann} J.~K.,  et~al., 2019, \mn@doi [\mnras] {10.1093/mnras/stz1539},
  \href {https://ui.adsabs.harvard.edu/abs/2019MNRAS.487.3650H} {487, 3650}

\bibitem[\protect\citeauthoryear{{Johansson} et~al.,}{{Johansson}
  et~al.}{2013}]{2013JohanssonJ}
{Johansson} J.,  et~al., 2013, \mn@doi [\mnras] {10.1093/mnras/stt1408}, \href
  {https://ui.adsabs.harvard.edu/abs/2013MNRAS.435.1680J} {435, 1680}

\bibitem[\protect\citeauthoryear{{Johansson} et~al.,}{{Johansson}
  et~al.}{2021}]{2021johansson}
{Johansson} J.,  et~al., 2021, \mn@doi [\apj] {10.3847/1538-4357/ac2f9e}, \href
  {https://ui.adsabs.harvard.edu/abs/2021ApJ...923..237J} {923, 237}

\bibitem[\protect\citeauthoryear{{Johnson}, {Leja}, {Conroy}  \&
  {Speagle}}{{Johnson} et~al.}{2021}]{2021_prospector}
{Johnson} B.~D.,  {Leja} J.,  {Conroy} C.,   {Speagle} J.~S.,  2021, \mn@doi
  [\apjs] {10.3847/1538-4365/abef67}, \href
  {https://ui.adsabs.harvard.edu/abs/2021ApJS..254...22J} {254, 22}

\bibitem[\protect\citeauthoryear{{Kelly}, {Hicken}, {Burke}, {Mandel}  \&
  {Kirshner}}{{Kelly} et~al.}{2010}]{kelly2010}
{Kelly} P.~L.,  {Hicken} M.,  {Burke} D.~L.,  {Mandel} K.~S.,   {Kirshner}
  R.~P.,  2010, \mn@doi [\apj] {10.1088/0004-637X/715/2/743}, \href
  {https://ui.adsabs.harvard.edu/abs/2010ApJ...715..743K} {715, 743}

\bibitem[\protect\citeauthoryear{{Kelsey} et~al.,}{{Kelsey}
  et~al.}{2021}]{2021Kelsey}
{Kelsey} L.,  et~al., 2021, \mn@doi [\mnras] {10.1093/mnras/staa3924}, \href
  {https://ui.adsabs.harvard.edu/abs/2021MNRAS.501.4861K} {501, 4861}

\bibitem[\protect\citeauthoryear{Kelsey et~al.,}{Kelsey
  et~al.}{2022}]{kelsey2022}
Kelsey L.,  et~al., 2022, Concerning Colour: The Effect of Environment on Type
  Ia Supernova Colour in the Dark Energy Survey,
  \mn@doi{10.48550/ARXIV.2208.01357}, \url {https://arxiv.org/abs/2208.01357}

\bibitem[\protect\citeauthoryear{{Kessler} \& {Scolnic}}{{Kessler} \&
  {Scolnic}}{2017}]{Kessler_2017}
{Kessler} R.,  {Scolnic} D.,  2017, \mn@doi [\apj]
  {10.3847/1538-4357/836/1/56}, \href
  {https://ui.adsabs.harvard.edu/abs/2017ApJ...836...56K} {836, 56}

\bibitem[\protect\citeauthoryear{{Khetan} et~al.,}{{Khetan}
  et~al.}{2021}]{Khetan2020}
{Khetan} N.,  et~al., 2021, \mn@doi [\aap] {10.1051/0004-6361/202039196}, \href
  {https://ui.adsabs.harvard.edu/abs/2021A&A...647A..72K} {647, A72}

\bibitem[\protect\citeauthoryear{{Lampeitl} et~al.,}{{Lampeitl}
  et~al.}{2010}]{Lampeitl_2010}
{Lampeitl} H.,  et~al., 2010, \mn@doi [\apj] {10.1088/0004-637X/722/1/566},
  \href {https://ui.adsabs.harvard.edu/abs/2010ApJ...722..566L} {722, 566}

\bibitem[\protect\citeauthoryear{{Lidman} et~al.,}{{Lidman}
  et~al.}{2020}]{Lidman_2020}
{Lidman} C.,  et~al., 2020, \mn@doi [\mnras] {10.1093/mnras/staa1341}, \href
  {https://ui.adsabs.harvard.edu/abs/2020MNRAS.496...19L} {496, 19}

\bibitem[\protect\citeauthoryear{Meldorf et~al.,}{Meldorf
  et~al.}{2022}]{meldorf2022}
Meldorf C.,  et~al., 2022, The Dark Energy Survey Supernova Program results:
  Type Ia Supernova brightness correlates with host galaxy dust,
  \mn@doi{10.48550/ARXIV.2206.06928}, \url {https://arxiv.org/abs/2206.06928}

\bibitem[\protect\citeauthoryear{{M{\"o}ller} \& {de
  Boissi{\`e}re}}{{M{\"o}ller} \& {de Boissi{\`e}re}}{2020}]{2020Anais}
{M{\"o}ller} A.,  {de Boissi{\`e}re} T.,  2020, \mn@doi [\mnras]
  {10.1093/mnras/stz3312}, \href
  {https://ui.adsabs.harvard.edu/abs/2020MNRAS.491.4277M} {491, 4277}

\bibitem[\protect\citeauthoryear{{M{\"o}ller} et~al.,}{{M{\"o}ller}
  et~al.}{2022}]{2022Anais}
{M{\"o}ller} A.,  et~al., 2022, The Dark Energy Survey 5-year photometrically
  identified Type Ia Supernovae (\mn@eprint {arXiv} {2112.02517})

\bibitem[\protect\citeauthoryear{{Morganson} et~al.,}{{Morganson}
  et~al.}{2018}]{Morganson_2018}
{Morganson} E.,  et~al., 2018, \mn@doi [\pasp] {10.1088/1538-3873/aab4ef},
  \href {https://ui.adsabs.harvard.edu/abs/2018PASP..130g4501M} {130, 074501}

\bibitem[\protect\citeauthoryear{{Orellana} et~al.,}{{Orellana}
  et~al.}{2017}]{2017Orellana}
{Orellana} G.,  et~al., 2017, \mn@doi [\aap] {10.1051/0004-6361/201629009},
  \href {https://ui.adsabs.harvard.edu/abs/2017A&A...602A..68O} {602, A68}

\bibitem[\protect\citeauthoryear{{Osterbrock}}{{Osterbrock}}{1989}]{1989osterbrock}
{Osterbrock} D.~E.,  1989, {Astrophysics of Gaseous Nebulae and Active Galactic
  Nuclei}.
University Science Books

\bibitem[\protect\citeauthoryear{{Pan} et~al.,}{{Pan} et~al.}{2014}]{Pan2013}
{Pan} Y.~C.,  et~al., 2014, \mn@doi [\mnras] {10.1093/mnras/stt2287}, \href
  {https://ui.adsabs.harvard.edu/abs/2014MNRAS.438.1391P} {438, 1391}

\bibitem[\protect\citeauthoryear{{Perlmutter} et~al.,}{{Perlmutter}
  et~al.}{1999}]{1999perl}
{Perlmutter} S.,  et~al., 1999, \mn@doi [\apj] {10.1086/307221}, \href
  {https://ui.adsabs.harvard.edu/abs/1999ApJ...517..565P} {517, 565}

\bibitem[\protect\citeauthoryear{{Phillips}}{{Phillips}}{1993}]{1993ApJ...413L.105P}
{Phillips} M.~M.,  1993, \mn@doi [\apjl] {10.1086/186970}, \href
  {https://ui.adsabs.harvard.edu/abs/1993ApJ...413L.105P} {413, L105}

\bibitem[\protect\citeauthoryear{{Pursiainen} et~al.,}{{Pursiainen}
  et~al.}{2020}]{Pursiainen2020}
{Pursiainen} M.,  et~al., 2020, \mn@doi [\mnras] {10.1093/mnras/staa995}, \href
  {https://ui.adsabs.harvard.edu/abs/2020MNRAS.494.5576P} {494, 5576}

\bibitem[\protect\citeauthoryear{{Riess} et~al.,}{{Riess}
  et~al.}{1998}]{Reiss1998}
{Riess} A.~G.,  et~al., 1998, \mn@doi [\aj] {10.1086/300499}, \href
  {http://adsabs.harvard.edu/abs/1998AJ....116.1009R} {116, 1009}

\bibitem[\protect\citeauthoryear{{Riess} et~al.,}{{Riess}
  et~al.}{2016}]{2016Riess}
{Riess} A.~G.,  et~al., 2016, \mn@doi [\apj] {10.3847/0004-637X/826/1/56},
  \href {https://ui.adsabs.harvard.edu/abs/2016ApJ...826...56R} {826, 56}

\bibitem[\protect\citeauthoryear{Riess et~al.,}{Riess et~al.}{2021}]{2022Riess}
Riess A.~G.,  et~al., 2021, arXiv e-prints, \href
  {https://arxiv.org/abs/2112.04510} {p. arXiv:2112.04510}

\bibitem[\protect\citeauthoryear{{Rigault} et~al.,}{{Rigault}
  et~al.}{2013}]{Rigault_2013}
{Rigault} M.,  et~al., 2013, \mn@doi [\aap] {10.1051/0004-6361/201322104},
  \href {https://ui.adsabs.harvard.edu/abs/2013A&A...560A..66R} {560, A66}

\bibitem[\protect\citeauthoryear{{Rigault} et~al.,}{{Rigault}
  et~al.}{2020}]{Rigault_2020}
{Rigault} M.,  et~al., 2020, \mn@doi [\aap] {10.1051/0004-6361/201730404},
  \href {https://ui.adsabs.harvard.edu/abs/2020A&A...644A.176R} {644, A176}

\bibitem[\protect\citeauthoryear{{Roman} et~al.,}{{Roman}
  et~al.}{2018}]{2018roman}
{Roman} M.,  et~al., 2018, \mn@doi [\aap] {10.1051/0004-6361/201731425}, \href
  {https://ui.adsabs.harvard.edu/abs/2018A&A...615A..68R} {615, A68}

\bibitem[\protect\citeauthoryear{{Rose}, {Garnavich}  \& {Berg}}{{Rose}
  et~al.}{2019}]{2019rose}
{Rose} B.~M.,  {Garnavich} P.~M.,   {Berg} M.~A.,  2019, \mn@doi [\apj]
  {10.3847/1538-4357/ab0704}, \href
  {https://ui-adsabs-harvard-edu.ezproxy.lib.swin.edu.au/abs/2019ApJ...874...32R}
  {874, 32}

\bibitem[\protect\citeauthoryear{{Salim}, {Boquien}  \& {Lee}}{{Salim}
  et~al.}{2018}]{2018Salim}
{Salim} S.,  {Boquien} M.,   {Lee} J.~C.,  2018, \mn@doi [\apj]
  {10.3847/1538-4357/aabf3c}, \href
  {https://ui.adsabs.harvard.edu/abs/2018ApJ...859...11S} {859, 11}

\bibitem[\protect\citeauthoryear{{Scolnic} \& {Kessler}}{{Scolnic} \&
  {Kessler}}{2016}]{2016Scolnic}
{Scolnic} D.,  {Kessler} R.,  2016, \mn@doi [\apjl]
  {10.3847/2041-8205/822/2/L35}, \href
  {https://ui.adsabs.harvard.edu/abs/2016ApJ...822L..35S} {822, L35}

\bibitem[\protect\citeauthoryear{{Scolnic} et~al.,}{{Scolnic}
  et~al.}{2018}]{2018Scolnic}
{Scolnic} D.,  et~al., 2018, \mn@doi [\apj] {10.3847/1538-4357/aab9bb}, \href
  {https://ui.adsabs.harvard.edu/abs/2018ApJ...859..101S} {859, 101}

\bibitem[\protect\citeauthoryear{{Smith} et~al.,}{{Smith}
  et~al.}{2020a}]{Smith2020b}
{Smith} M.,  et~al., 2020a, \mn@doi [\aj] {10.3847/1538-3881/abc01b}, \href
  {https://ui.adsabs.harvard.edu/abs/2020AJ....160..267S} {160, 267}

\bibitem[\protect\citeauthoryear{{Smith} et~al.,}{{Smith}
  et~al.}{2020b}]{Smith_2020}
{Smith} M.,  et~al., 2020b, \mn@doi [\mnras] {10.1093/mnras/staa946}, \href
  {https://ui.adsabs.harvard.edu/abs/2020MNRAS.494.4426S} {494, 4426}

\bibitem[\protect\citeauthoryear{Sullivan et~al.,}{Sullivan
  et~al.}{2010}]{Sullivan_2010}
Sullivan M.,  et~al., 2010, \mn@doi [\mnras] {10.1111/j.1365-2966.2010.16731.x}

\bibitem[\protect\citeauthoryear{{Swann} et~al.,}{{Swann}
  et~al.}{2019}]{2019Swann}
{Swann} E.,  et~al., 2019, \mn@doi [arXiv: Instrumentation and Methods for
  Astrophysics] {10.18727/0722-6691/5129}

\bibitem[\protect\citeauthoryear{{Triani}, {Sinha}, {Croton}, {Dwek}  \&
  {Pacifici}}{{Triani} et~al.}{2021}]{2021Triani}
{Triani} D.~P.,  {Sinha} M.,  {Croton} D.~J.,  {Dwek} E.,   {Pacifici} C.,
  2021, \mn@doi [\mnras] {10.1093/mnras/stab558}, \href
  {https://ui.adsabs.harvard.edu/abs/2021MNRAS.503.1005T} {503, 1005}

\bibitem[\protect\citeauthoryear{{Tripp}}{{Tripp}}{1998}]{1998A&A...331..815T}
{Tripp} R.,  1998, \aap, \href
  {https://ui.adsabs.harvard.edu/abs/1998A&A...331..815T} {331, 815}

\bibitem[\protect\citeauthoryear{{Uddin}, {Mould}, {Lidman}, {Ruhlmann-Kleider}
   \& {Hardin}}{{Uddin}
  et~al.}{2017}]{uddin_mould_lidman_ruhlmann-kleider_hardin_2017}
{Uddin} S.~A.,  {Mould} J.,  {Lidman} C.,  {Ruhlmann-Kleider} V.,   {Hardin}
  D.,  2017, \mn@doi [\pasa] {10.1017/pasa.2017.2}, \href
  {https://ui.adsabs.harvard.edu/abs/2017PASA...34....9U} {34, e009}

\bibitem[\protect\citeauthoryear{{Vazdekis}, {S{\'a}nchez-Bl{\'a}zquez},
  {Falc{\'o}n-Barroso}, {Cenarro}, {Beasley}, {Cardiel}, {Gorgas}  \&
  {Peletier}}{{Vazdekis} et~al.}{2010}]{miles}
{Vazdekis} A.,  {S{\'a}nchez-Bl{\'a}zquez} P.,  {Falc{\'o}n-Barroso} J.,
  {Cenarro} A.~J.,  {Beasley} M.~A.,  {Cardiel} N.,  {Gorgas} J.,   {Peletier}
  R.~F.,  2010, \mn@doi [\mnras] {10.1111/j.1365-2966.2010.16407.x}, \href
  {https://ui.adsabs.harvard.edu/abs/2010MNRAS.404.1639V} {404, 1639}

\bibitem[\protect\citeauthoryear{{Wiseman} et~al.,}{{Wiseman}
  et~al.}{2020a}]{Wiseman_2020}
{Wiseman} P.,  et~al., 2020a, \mn@doi [\mnras] {10.1093/mnras/staa1302}, \href
  {https://ui.adsabs.harvard.edu/abs/2020MNRAS.495.4040W} {495, 4040}

\bibitem[\protect\citeauthoryear{{Wiseman} et~al.,}{{Wiseman}
  et~al.}{2020b}]{Wiseman2020b}
{Wiseman} P.,  et~al., 2020b, \mn@doi [\mnras] {10.1093/mnras/staa2474}, \href
  {https://ui.adsabs.harvard.edu/abs/2020MNRAS.498.2575W} {498, 2575}

\bibitem[\protect\citeauthoryear{{Yuan} et~al.,}{{Yuan}
  et~al.}{2015}]{Yuan2015}
{Yuan} F.,  et~al., 2015, \mn@doi [\mnras] {10.1093/mnras/stv1507}, \href
  {http://adsabs.harvard.edu/abs/2015MNRAS.452.3047Y} {452, 3047}

\makeatother
\end{thebibliography}




\appendix

\section{Author Affiliations}
$^{1}$ Centre for Astrophysics \& Supercomputing, Swinburne University of Technology, Victoria 3122, Australia \\
$^{2}$ Centre for Gravitational Astrophysics, College of Science, The Australian National University, ACT 2601, Australia \\
$^{3}$ The Research School of Astronomy and Astrophysics, Australian National University, ACT 2601, Australia \\
$^{4}$ Institute of Cosmology and Gravitation, University of Portsmouth, Portsmouth, PO1 3FX, UK \\
$^{5}$ School of Physics and Astronomy, University of Southampton,  Southampton, SO17 1BJ, UK \\
$^{6}$ Center for Astrophysics $\vert$ Harvard \& Smithsonian, 60 Garden Street, Cambridge, MA 02138, USA \\
$^{7}$ Institut d'Estudis Espacials de Catalunya (IEEC), 08034 Barcelona, Spain \\
$^{8}$ Institute of Space Sciences (ICE, CSIC),  Campus UAB, Carrer de Can Magrans, s/n,  08193 Barcelona, Spain \\
$^{9}$ School of Mathematics and Physics, University of Queensland,  Brisbane, QLD 4072, Australia \\
$^{10}$ Department of Physics, Duke University Durham, NC 27708, USA \\
$^{11}$ Sydney Institute for Astronomy, School of Physics, A28, The University of Sydney, NSW 2006, Australia \\
$^{12}$ Department of Astronomy and Astrophysics, University of Chicago, Chicago, IL 60637, USA \\
$^{13}$ Kavli Institute for Cosmological Physics, University of Chicago, Chicago, IL 60637, USA \\
$^{14}$ Cerro Tololo Inter-American Observatory, NSF's National Optical-Infrared Astronomy Research Laboratory, Casilla 603, La Serena, Chile \\
$^{15}$ Laborat\'orio Interinstitucional de e-Astronomia - LIneA, Rua Gal. Jos\'e Cristino 77, Rio de Janeiro, RJ - 20921-400, Brazil \\
$^{16}$ Fermi National Accelerator Laboratory, P. O. Box 500, Batavia, IL 60510, USA \\
$^{17}$ Department of Physics, University of Michigan, Ann Arbor, MI 48109, USA \\
$^{18}$ Centro de Investigaciones Energ\'eticas, Medioambientales y Tecnol\'ogicas (CIEMAT), Madrid, Spain \\
$^{19}$ CNRS, UMR 7095, Institut d'Astrophysique de Paris, F-75014, Paris, France \\
$^{20}$ Sorbonne Universit\'es, UPMC Univ Paris 06, UMR 7095, Institut d'Astrophysique de Paris, F-75014, Paris, France \\
$^{21}$ University Observatory, Faculty of Physics, Ludwig-Maximilians-Universit\"at, Scheinerstr. 1, 81679 Munich, Germany \\
$^{22}$ Department of Physics \& Astronomy, University College London, Gower Street, London, WC1E 6BT, UK \\
$^{23}$ Kavli Institute for Particle Astrophysics \& Cosmology, P. O. Box 2450, Stanford University, Stanford, CA 94305, USA \\
$^{24}$ SLAC National Accelerator Laboratory, Menlo Park, CA 94025, USA \\
$^{25}$ Instituto de Astrofisica de Canarias, E-38205 La Laguna, Tenerife, Spain \\
$^{26}$ Universidad de La Laguna, Dpto. Astrofísica, E-38206 La Laguna, Tenerife, Spain \\
$^{27}$ INAF-Osservatorio Astronomico di Trieste, via G. B. Tiepolo 11, I-34143 Trieste, Italy \\
$^{28}$ Center for Astrophysical Surveys, National Center for Supercomputing Applications, 1205 West Clark St., Urbana, IL 61801, USA \\
$^{29}$ Department of Astronomy, University of Illinois at Urbana-Champaign, 1002 W. Green Street, Urbana, IL 61801, USA \\
$^{30}$ Institut de F\'{\i}sica d'Altes Energies (IFAE), The Barcelona Institute of Science and Technology, Campus UAB, 08193 Bellaterra (Barcelona) Spain \\
$^{31}$ Astronomy Unit, Department of Physics, University of Trieste, via Tiepolo 11, I-34131 Trieste, Italy \\
$^{32}$ Institute for Fundamental Physics of the Universe, Via Beirut 2, 34014 Trieste, Italy \\
$^{33}$ Hamburger Sternwarte, Universit\"{a}t Hamburg, Gojenbergsweg 112, 21029 Hamburg, Germany \\
$^{34}$ Jet Propulsion Laboratory, California Institute of Technology, 4800 Oak Grove Dr., Pasadena, CA 91109, USA \\
$^{35}$ Institute of Theoretical Astrophysics, University of Oslo. P.O. Box 1029 Blindern, NO-0315 Oslo, Norway \\
$^{36}$ Instituto de Fisica Teorica UAM/CSIC, Universidad Autonoma de Madrid, 28049 Madrid, Spain \\
$^{37}$ Department of Physics and Astronomy, University of Pennsylvania, Philadelphia, PA 19104, USA \\
$^{38}$ Department of Astronomy, University of Michigan, Ann Arbor, MI 48109, USA \\
$^{39}$ Observat\'orio Nacional, Rua Gal. Jos\'e Cristino 77, Rio de Janeiro, RJ - 20921-400, Brazil \\
$^{40}$ Santa Cruz Institute for Particle Physics, Santa Cruz, CA 95064, USA \\
$^{41}$ Center for Cosmology and Astro-Particle Physics, The Ohio State University, Columbus, OH 43210, USA \\
$^{42}$ Department of Physics, The Ohio State University, Columbus, OH 43210, USA \\
$^{43}$ Australian Astronomical Optics, Macquarie University, North Ryde, NSW 2113, Australia \\
$^{44}$ Lowell Observatory, 1400 Mars Hill Rd, Flagstaff, AZ 86001, USA \\
$^{45}$ Instituci\'o Catalana de Recerca i Estudis Avan\c{c}ats, E-08010 Barcelona, Spain \\
$^{46}$ Physics Department, 2320 Chamberlin Hall, University of Wisconsin-Madison, 1150 University Avenue Madison, WI  53706-1390 \\
$^{47}$ Department of Physics, University of Surrey, Guilford, Surrey, UK \\
$^{48}$ Department of Astronomy, University of California, Berkeley,  501 Campbell Hall, Berkeley, CA 94720, USA \\
$^{49}$ Institute of Astronomy, University of Cambridge, Madingley Road, Cambridge CB3 0HA, UK \\
$^{50}$ Department of Astrophysical Sciences, Princeton University, Peyton Hall, Princeton, NJ 08544, USA \\
$^{51}$ Department of Physics and Astronomy, Pevensey Building, University of Sussex, Brighton, BN1 9QH, UK \\
$^{52}$ Computer Science and Mathematics Division, Oak Ridge National Laboratory, Oak Ridge, TN 37831 \\
$^{53}$ Excellence Cluster Origins, Boltzmannstr.\ 2, 85748 Garching, Germany \\
$^{54}$ Max Planck Institute for Extraterrestrial Physics, Giessenbachstrasse, 85748 Garching, Germany \\
$^{55}$ Universit\"ats-Sternwarte, Fakult\"at f\"ur Physik, Ludwig-Maximilians Universit\"at M\"unchen, Scheinerstr. 1, 81679 M\"unchen, Germany \\


\bsp	
\label{lastpage}
\end{document}